\RequirePackage{snapshot}
\documentclass[%
reprint,
amsfonts,
floatfix,
amsmath,amssymb,
aps,
pra,
]{revtex4-2}
\usepackage[version=4]{mhchem}
\usepackage{graphicx}
\usepackage{epstopdf}
\usepackage{dcolumn}
\usepackage{bm}
\usepackage{todonotes}
\usepackage{xargs}
\usepackage{xcolor}
\usepackage{siunitx}
\usepackage{caption}
\usepackage{booktabs}
\usepackage{csquotes}
\usepackage{letltxmacro}
\usepackage{pgfplots}
\usepackage[labelformat=simple]{subcaption}
\usepackage{float}
\usepackage{stackengine}
\usepackage{tikz}
\usepackage{tikzscale}
\usetikzlibrary{external}
\newcommand{\TikzCachePath}{tikz-cache/}
\tikzset{external/optimize=false}
\usetikzlibrary{pgfplots.groupplots}

\graphicspath{{./graphs/tikz-pictures}}

\ifluatex
\usepackage{pdftexcmds}
\makeatletter
\let\pdfstrcmp\pdf@strcmp
\let\pdffilemoddate\pdf@filemoddate
\makeatother
\else
\fi

\makeatletter
\renewcommand{\todo}[2][]{%
	\begingroup \tikzexternaldisable
	\@todo[#1]{#2} \tikzexternalenable
	\endgroup}
\makeatother

\makeatletter
\newcommand{\mathcolor}[2]{%
	\begingroup
	\colorlet{out}{.}\color{#1}#2\@ifnextchar_{\do@mathcolorsub}{\endgroup}%
}
\newcommand{\do@mathcolorsub}[2]{%
	_{{\color{out}#2}}\@ifnextchar^{\do@mathcolorsup}{\endgroup}%
}
\newcommand{\do@mathcolorsup}[2]{%
	^{\color{out}#2}\endgroup
}
\makeatother

\newcommandx{\unsure}[2][1=]{\tikzexternaldisable\todo[linecolor=red,backgroundcolor=red!25,bordercolor=red,#1]{#2}\tikzexternalenable}
\newcommandx{\warningtodo}[2][1=]{\tikzexternaldisable\todo[linecolor=red,backgroundcolor=red,bordercolor=red, textcolor=white,#1]{\Huge #2}\tikzexternalenable}
\newcommandx{\change}[2][1=]{\tikzexternaldisable\todo[linecolor=blue,backgroundcolor=blue!25,bordercolor=blue,#1]{#2}\tikzexternalenable}
\newcommandx{\info}[2][1=]{\tikzexternaldisable\todo[linecolor=OliveGreen,backgroundcolor=OliveGreen!25,bordercolor=OliveGreen,#1]{#2}\tikzexternalenable}
\newcommandx{\improvement}[2][1=]{\tikzexternaldisable\todo[linecolor=Plum,backgroundcolor=Plum!25,bordercolor=Plum,#1]{#2}\tikzexternalenable}
\newcommandx{\thiswillnotshow}[2][1=]{\tikzexternaldisable\todo[disable,#1]{#2}\tikzexternalenable}

\newcommand{\figurefontsize}{\footnotesize}
\pgfplotsset{every axis/.append style={
		label style={font=\figurefontsize\bfseries},
		tick label style={font=\figurefontsize},
		legend style={font=\figurefontsize},
		every node/.style={font=\figurefontsize}  
	},
	y axis/.append style={align=center}}
\tikzset{Line Label/.style={font=\figurefontsize,scale=2}}

\DeclareMathOperator{\sech}{sech}

\newcommand{\actualquote}[1]{\enquote{#1}}
\newcommand{\definedterm}[1]{\enquote{#1}}

\usepackage{xparse}
\NewDocumentCommand{\framecolorbox}{oommm}
{
	\IfValueTF{#1}
	{%
		\IfValueTF{#2}
		{\fcolorbox{#3}{#4}{\makebox[#1][#2]{#5}}}
		{\fcolorbox{#3}{#4}{\makebox[#1]{#5}}}%
	}
	{\fcolorbox{#3}{#4}{#5}}%
}

\usepackage{hyperref}
\usepackage{xcolor}
\hypersetup{
	colorlinks,
	linkcolor={red!50!black},
	citecolor={blue!50!black},
	urlcolor={blue!80!black}
}

\usepackage{cleveref}
\pgfplotsset{compat=newest}

\newcommand{\leftspecialcell}[2][c]{%
	\begin{tabular}[#1]{@{}l@{}}#2\end{tabular}}

\newcommand{\NFOMFigWidth}{0.8}
\newcommand{\NFOMFigHeigth}{0.2}

\makeatletter

\newcommand{\OverLeafFigures}{}

\AtBeginDocument{%
	\LetLtxMacro{\includegraphicsorig}{\includegraphics}
	\RenewDocumentCommand{\includegraphics}{O{} m }{%
		\filename@parse{#2}
		\IfStrEq{\detokenize\expandafter{\filename@ext}}{\detokenize{tikz}}{%
			\ifdefined\OverLeafFigures%
			\includegraphicsorig[#1]{\TikzCachePath\filename@base.pdf}%
			\else%
			\tikzsetnextfilename{\filename@base}%
			\includegraphicsorig[#1]{\filename@area\filename@base.tikz}%
			\fi%
		}{%
			\includegraphicsorig[#1]{#2}%
		}%
	}%
}%
\makeatother

\begin{document}
	
	\preprint{APS/123-QED}
	
	\title{Nonlinearity and wavelength control in ultrashort-pulse subsurface material processing}
	
	\author{Roland~A.~Richter}
	\email{roland.richter@ntnu.no}
	\affiliation{Department of Physics, NTNU, Norwegian University of Science and Technology, Trondheim, Norway}
	\author{Vladimir~Kalashnikov}%
	\affiliation{Photonics Institute, Vienna University of Technology, Vienna, Austria}
	\affiliation{Dipartimento di Ingegneria dell'Informazione, Sapienza University of Rome, Italy}
	\author{Irina~T.~Sorokina}
	\affiliation{Department of Physics, NTNU, Norwegian University of Science and Technology, Trondheim, Norway}%
	\affiliation{ATLA Lasers AS, Trondheim, Norway}
	
	\date{\today}
	
	\begin{abstract}
		The pronounced dependence of the nonlinear parameters of both dielectric and semiconductor materials on the wavelength, and the nonlinear interaction between the ultra-short laser pulse and the material requires precise control of the wavelength of the pulse, in addition to the precise control of the pulse energy, pulse duration and focusing optics. This becomes particularly important for fine sub-wavelength single pulse sub-surface processing. Based on two different numerical models and taking \ce{Si} as example material, we investigate the spatio-temporal behavior of a pulse propagating through the material while covering a broad range of parameters. The wavelength-dependence of material processing depends on the different contributions of two- and tree-photon absorption in combination with the Kerr effect which results in a particularly sharp nonlinear peak at $ \approx\SI{2100}{\nano\meter} $. We could show that in silicon this makes processing preferable close to this wavelength. The impact of the nonlinear nonparaxial propagation effects on spatio-temporal beam structure is also investigated. It could be shown that with increasing wavelength and large focusing angles the aberrations at the focal spot can be reduced, and thereby cleaner and more precise processing can be achieved. Finally, we could show that the optimum energy transfer from the pulse to the material is within a narrow window of pulse durations between \SIrange[range-units=single, range-phrase={ to }]{600}{900}{\femto\second}.
	\end{abstract}
	
	\maketitle
	
	\section{Introduction \& State of the Art}
	\label{sec:Intro}
	Fine material processing of semiconductors in general, and silicon in particular is a hot topic both in the scientific community as well as in industry. Up to now a lot of effort has been put into surface processing experiments using mainly nanosecond pulse durations and pulse energies above \SI{1}{\micro\joule}~\cite{Ohmura2006,Kumagai2006,Ohmura2007,Chambonneau2016,Franta2018,Chambonneau2018,Chambonneau2021,Panchal2021}. Processing technology has been well understood both on the practical and theoretical level~\cite{Verburg,Chambonneau2021}. In the last few years the focus has been shifted towards the next level fine material processing -- in particular sub-surface modification of the crystalline silicon using ultra-short pulsed lasers (pico- and femtosecond) at the wavelengths above \SI{1100}{\nano\meter}, where silicon becomes transparent. Several successful studies have been carried out since then~\cite{Nejadmalayeri2005,Zavedeev2014,Grojo2015,Mori2015,Shimotsuma2016,Zavedeev2016,Richter2018,Kammer2018,Das2020,Mareev2020,Blothe2021}. Those sub-surface defects are targeted for such important industrial processes as thin wafer exfoliation~\cite{Richter2020}, sub-surface structuring~\cite{Kammer2018,Richter2020}, data storage~\cite{Tokel2017}, waveguides~\cite{Chambonneau2016,Tokel2017,Pavlov2017,Turnali:19} and grating fabrication~\cite{Chambonneau2018}. The common way to introduce sub-surface modifications in all these works is using multiple pulses and pulse trains~\cite{Wang2020THz}. Very recently it was shown that also single pulse processing of monocrystalline silicon is possible with ultra-short pulses at \SI{2100}{\nano\meter}, producing particularly fine sub-wavelength modifications~\cite{Tolstik2022CLEO}. Besides optimization of the laser pulse parameters, the choice of the laser wavelength played a crucial role in this work, suggesting a much more complex nonlinear light-matter interaction with spatio-temporal dynamics that strongly depend on the wavelength. The latter is a subject of the present numerical investigation.
	
	Besides nonlinear multi-photon absorption light propagating through the material experiences several different processes, such as self-focusing~\cite{Richter2018,Richter2020} and plasma-induced de-focusing and reflection~\cite{Zavedeev2014,Zavedeev2016}, to name a few. Those processes strongly depend on the wavelength and intensity of the pulse propagating through the material~\cite{Lin2007,Wang2013}. Depending on the wavelength the impact of each parameter will vary. Here, especially the self-focusing effect due to the third-order nonlinearity has to be mentioned, as it becomes extremely strong within a rather narrow wavelength range at a certain wavelength, that depends on the forbidden band-gap of the material. An investigation of the wavelength-dependent interplay of several nonlinear optical effects and their impact on the spatio-temporal dynamics of ultra-short laser-pulses inside the material becomes particularly important for single-pulse sub-surface processing of any material, not only silicon. In the present work we for the for the first time perform such study using silicon as a model. 
	
	In silicon the third-order nonlinearity peaks between \SIrange[range-units=single, range-phrase=\text{ to }]{2000}{2200}{\nano\meter}~\cite{Lin2007,Wang2013}, and therefore influences the pulse significantly stronger than at shorter or longer wavelengths. Within the same wavelength range the two- and three-photon absorption experiences a minimum at this wavelength, thus allowing to reduce the potential for surface damage (due to lower multi-photon absorption) while retaining the capability of delivering a large amount of energy to the focal spot~\cite{Richter2018}. This makes it possible to achieve pulse beam collapse and correspondingly high intensities at relatively low pulse energies. For certain parameters this might decrease the material processing threshold, while simultaneously increase the precision at which the material can be processed. In Ref.~\cite{Richter2020} we for the first time suggested to use the interplay of all these nonlinear optical effects to find an optimum wavelength for processing of silicon with \si{fs}- and \si{ps}-pulses. The modelling and experimental results were obtained at low NA values of \numrange{0.125}{0.3}, and it became clear that using a large focusing angle is necessary to avoid nonlinear absorption too close to the surface, as also mentioned by~\citet{Zavedeev2016}.
	
	In earlier theoretical works~\citet{Zavedeev2014,Zavedeev2016} it was shown that for a numerical aperture of \num{0.3} and a pulse duration of \SI{250}{\femto\second} the maximum deposited energy density reaches its maximum value at around \SI{2000}{\nano\meter} and then, after dropping down, starts to grow with the wavelength again. This is in contrast with the maximum plasma density, which decreases with $ \lambda $ nearly monotonically. Moreover,~\citet{Kononenko2016} showed that an increase of the pulse energy not necessarily will increase the intensity at the focal spot due to multi-photon absorption, plasma defocusing and reflection. All this suggested the need of careful control of the laser parameters, including wavelength, to make use of the strong wavelength dependence of the material nonlinearity, to achieve a particularly precise single pulse shot processing.
	
	Besides the wavelength, energy and pulse duration are equally important, as experimentally shown by~\citet{Das2020, Mareev2020}. They demonstrate that a certain pulse duration is necessary to create sub-surface modifications within \ce{Si} depending on the focusing angle.~\Citet{Das2020} gives a minimum pulse duration of \SI{5.4}{\pico\second} for a focusing numerical aperture of \num{0.85} to induce modifications within \ce{Si} when operating a laser at \SI{1550}{\nano\meter}. For pulses between \SIrange[range-units=single, range-phrase=\text{ to }]{5.4}{21}{\pico\second} no increase of the modification threshold could be found~\cite{Das2020}, except when increasing the pulse duration to nanoseconds. For shorter pulse durations in the femtosecond range~\citet{Mareev2020} could show a decreasing amount of generated free carriers~\cite{Mareev2020} when keeping the wavelength, pulse energy and focusing optics constant, indicating higher losses while propagating through \ce{Si} for shorter pulse durations. This was experimentally confirmed by~\citet{Chambonneau2020}.
	
	To the best of our knowledge~\citet{Zavedeev2016} was the first to have investigated numerically the behavior of femto-second pulses propagating through \ce{Si} at wavelengths $ \gg\SI{2000}{\nano\meter} $. They stated that the pre-focal depletion of the pulse energy decreases significantly with wavelength, and that the maximum deposited energy density and the plasma defocusing decreases. The decrease in the free-carrier concentration does not cause a similar decrease in the density of the deposited energy, because the inverse bremsstrahlung absorption becomes stronger with longer wavelengths. In spite of all these advantages of going to longer wavelengths they did not expect a significant improvement in femto-second laser processing when going above \SI{5}{\micro\meter} wavelength at an NA of \num{0.3}. 
	
	In the last few years large progress has been achieved in development of ultrashort pulsed high energy lasers operating above \SI{2000}{\nano\meter}. Three laser systems could be of potential interest in this wavelength region: a picosecond-fiber \ce{Ho}:MOPA laser, operating at up to \SI{10}{\milli\joule} at \SI{10}{\kilo\hertz} repetition rates~\cite{Richter2020}, a femtosecond \ce{Cr{:}ZnS}~\cite{Sorokina2015} or \ce{Cr{:}ZnSe} laser with up to \SI{2}{\milli\joule} pulse energy~\cite{Slobodchikov2016,Ren2018,Vasilyev2019}, and an \ce{Er}:ZBLAN fiber laser operating around \SI{3}{\micro\meter}, delivering pulse durations down to \SIrange[range-units=single, range-phrase=\text{ to }]{100}{200}{\femto\second}~\cite{Duval2015,Gu2020,Wang2020}. The availability of commercial lasers in this wavelength range calls for a detailed numerical analysis that might be particularly important for practical applications of these lasers.
	
	\subsection{Mechanisms of sub-surface modifications in silicon}
	\label{subsec:MechanismOfSubSurfaceModificationSiTheory}
	
	Several mechanism for the modification of the crystal structure of silicon under irradiation with short and ultra-short pulses have already been described. Depending on the pulse parameters and the chosen model each approach relies on different mechanisms, time scales and different thresholds. Below the most relevant approaches for the investigated pulse parameter range are given:
	\begin{enumerate}
		\item\label{item:ThresholdCold} So-called \actualquote{non-thermal melting} is proposed when irradiating silicon with ultra-short pulses~\cite{Biswas1982,Stampfli1990,Stampfli1992,Sokolowski-Tinten1995,Sokolowski-Tinten2000,Apostolova2020,Apostolova2022}. By exciting more than $ \approx\SIrange[range-units=single, range-phrase=\text{ to }]{5}{10}{\percent} $ of the valence electrons in the conductance band, corresponding to a plasma density of $ \approx\SIrange[range-units=single, range-phrase=\text{ to }]{5e21}{10e21}{\per\cubic\centi\meter} $) the bonds between the atoms are weakened enough to enable a process which resembles melting, but without reaching the lattice temperatures associated with melting. This process is independent of the wavelength, after it only relies on the density of excited electrons.
		\item\label{item:TresholdCritical} Following~\Citet{Du1994,B.C1995,Du1996,Tien1999,Chimier2011} reaching the critical plasma density in a material, defined as $ n_c=\left[\left(\varepsilon_0\cdot m_e\right)/e^2\right]\cdot\omega^2 $ is sufficient for reaching the modification/damage threshold in a material. This criterion is wavelength-dependent, and scales with $ \lambda^{-2} $. Therefore, it is easier to reach the modification threshold with this definition using longer wavelengths compared to shorter wavelengths.
		\item Following~\citet{Chimier2011,Zhokhov2018,Werner2019,Chambonneau2021} reaching the threshold given in point~\ref{item:TresholdCritical} is necessary for modifying the material, but not necessarily sufficient. Instead, they propose to relate to the deposited energy within the material when determining if the material will be modified. For ultra-short pulses it can be assumed that the energy initially is retained only within the electrons during the full pulse duration, and only afterwards will be transferred into the lattice. Therefore, as upper limit for the absorbed energy \Cref{equ:AbsorbedEnergy} can be used.
		\begin{equation}\label{equ:AbsorbedEnergy}
			E_\text{max}=2E_g\varrho_\text{nt}.
		\end{equation}
		Here, $ E_g $ gives the band gap, and $ \varrho_\text{nt} $ the generated free-carrier density. Other approaches for determining the threshold (such as the Gamaly-model~\cite{Gamaly2002}) are described by~\Citet{Werner2019}.
	\end{enumerate}
	\subsection{Applied methods and procedure}
	\label{subsec:AppliedMethodsAndProcedure}
	
	In this work we focus on the theoretical study of the spatio-temporal effects of the propagation of the ultra-short pulses inside silicon at different wavelengths above \SI{1100}{\nano\meter}. These effects are a result of the interplay of nonlinear optical effects and have a strong wavelength dependence. 
	
	To numerically investigate numerical apertures above \num{0.2} a non-paraxial pulse propagation method has to be used, the paraxial approximation is not sufficient for numerical apertures larger than that~\cite{Fedorov2016}. For this purpose two different methods are used in this paper, the unidirectional pulse propagation~\cite{Kolesik2002} and the generalized Helmholtz equation~\cite{PhysRevLett.76.4356}. While the former approach is a spectral technique using the Fourier split-step method for time stepping, the latter is a purely temporal technique, and is implemented and solved with a finite-element-method (FEM). This allows the generalized Helmholtz approach to better resolve spatial effects at the cost of neglecting spectral effects.
	
	We use numerical simulations to investigate the influence of the wavelength dependence of nonlinear absorption, self-focusing and plasma-related effects for wavelengths between \SIrange[range-units=single, range-phrase=\text{ to }]{1550}{3250}{\nano\meter} for pulse durations between \SIrange[range-units=single, range-phrase=\text{ to }]{100}{5000}{\femto\second} and compare the influence of the numerical aperture. Moreover, we also investigate the spatio-temporal behaviour of the pulse and demonstrate the significant influence of the generated plasma on the pulse shape while propagating through the material, and at the focal spot.
	
	We apply the methods of the spatio-temporal structure and dynamics control in the nonlinear media under the condition of strong field localization. Our observations demonstrate a possibility of spatio-temporal control like the nonlinear mode-self-cleaning in the multimode fibers~\cite{Krupa2017,Krupa2019} that bridges nonlinear fiber- and solid-state laser photonics. Manipulation with highly concentrated spatio-temporal patterns provides the tools for high-energy physics and technology in photonics, information processing, condensed matter modification and others.
	
	\section{Numerical model and physical parameters}
	\label{sec:NumModelAndPhysParams}
	\subsection{Physical nonlinear parameters}
	\label{subsec:PhysNonlinParams}
	A light pulse propagating through Silicon is influenced by several different wavelength-dependent nonlinear phenomena, such as the Kerr-effect, multi-photon absorption or plasma absorption and -defocusing.
	
\begin{figure*}[htb]
		\centering
		\tikzsetnextfilename{stacked_nonlinear_parameters}
		\includegraphics[width=0.9\linewidth, height=0.4\linewidth]{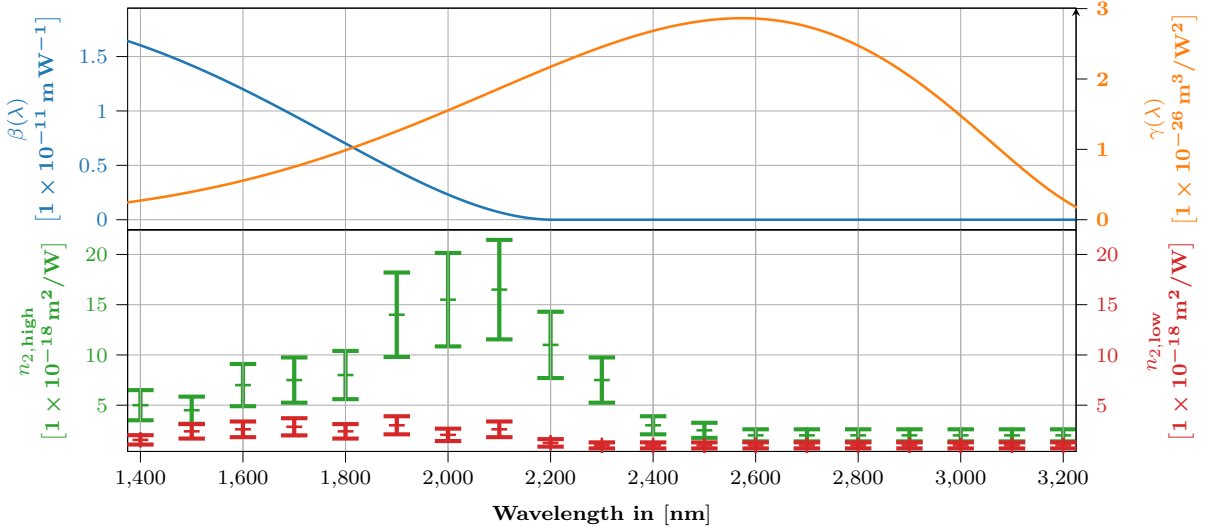}
		\caption{The values for the nonlinear parameters in \ce{Si} for a two-photon absorption $ \beta(\lambda) $ (the blue curve)~\cite{Bristow2007}, and a three-photon absorption $ \gamma(\lambda) $ (the orange curve)~\cite{Pearl2008} are shown in the upper figure. The highest and lowest $ n_2 $-values for the Kerr-coefficient (the green and red error bars, respectively) are shown in the lower part of the figure~\cite{Wang2013,Lin2007}. The curve for four-photon absorption (based on \citet{Gai2013}) is not shown here due to the limitation to \SI{3250}{\nano\meter} as the upper wavelength range.}
		\label{fig:nPA-Absorption}
	\end{figure*}

	Within the investigated wavelength range (between \SI{1100}{\nano\meter} and \SI{2500}{\nano\meter}) only two- and three-photon absorption can be seen as relevant, after single-photon absorption can be neglected for wavelengths above \SI{1100}{\nano\meter}, and four-photon absorption can be neglected for wavelengths below $ \approx\SI{3300}{\nano\meter} $. A graph displaying the values for the two-photon absorption (based on~\citet{Bristow2007} is shown in \Cref{fig:nPA-Absorption} in the upper figure in blue, and the values for the three-photon absorption (based on~\citet{Pearl2008}) are shown in the same figure in orange.
	
	The exact values for the wavelengths (as given in \Cref{sec:NumModelAndPhysParams}) investigated in this paper are shown in \Cref{tab:PhysParamsExample}, together with the corresponding sources.
	
	\begin{table*}[htb]
		\centering
		\caption{Physical parameter values for nonlinear interactions acting on light propagating through \ce{Si} at the example wavelengths. The data is based on \Cref{fig:nPA-Absorption} with the corresponding literature.}
		\label{tab:PhysParamsExample}
		\begin{tabular}{llllll}
			\toprule
			Wavelength&\leftspecialcell{Two-photon\\absorption\\coefficient~\cite{Bristow2007}}&\leftspecialcell{Three-photon\\absorption\\coefficient~\cite{Pearl2008}}&\leftspecialcell{Four-photon\\absorption\\coefficient~\cite{Gai2013}}&\leftspecialcell{Nonlinear refrac-\\tive index~\cite{Wang2013}}&\leftspecialcell{Refractive\\index~\cite{Li1980}}\\
			\midrule
			\SI{1550}{\nano\meter}&\SI{1.313e-11}{\meter\per\watt}&\SI{4.711e-27}{\meter\cubed\per\watt\squared}&\SI{0}{\meter^5\per\watt\cubed}&\SI{5.75e-18}{\meter\squared\per\watt}&\num{3.4757}\\
			\SI{1950}{\nano\meter}&\SI{3.361e-12}{\meter\per\watt}&\SI{1.403e-26}{\meter\cubed\per\watt\squared}&\SI{0}{\meter^5\per\watt\cubed}&\SI{1.475e-17}{\meter\squared\per\watt}&\num{3.453}\\
			\SI{2150}{\nano\meter}&\SI{2.030e-13}{\meter\per\watt}&\SI{2.020e-26}{\meter\cubed\per\watt\squared}&\SI{0}{\meter^5\per\watt\cubed}&\SI{1.375e-17}{\meter\squared\per\watt}&\num{3.446}\\
			\SI{2550}{\nano\meter}&\SI{0}{\meter\per\watt}&\SI{2.860e-26}{\meter\cubed\per\watt\squared}&\SI{0}{\meter^5\per\watt\cubed}&\SI{2.25e-18}{\meter\squared\per\watt}&\num{3.437}\\
			\SI{3250}{\nano\meter}&\SI{0}{\meter\per\watt}&\SI{9.366e-28}{\meter\cubed\per\watt\squared}&\SI{1e-41}{\meter^5\per\watt\cubed}&\SI{2e-18}{\meter\squared\per\watt}&\num{3.428}\\
			\bottomrule
		\end{tabular}
	\end{table*}
	
	The strength of the Kerr-effect can vary depending on the doping and orientation of the silicon wafer, and therefore we chose to include two different values as an upper and lower boundary. The wavelength-dependent values are shown in the lower figure in \Cref{fig:nPA-Absorption} for low $ n_2 $-values (based on~\citet{Lin2007}) in red and in green for high $ n_2 $-values (based on~\citet{Wang2013}). The main difference between the used materials in~\citet{Lin2007,Wang2013} is the different doping of the material, which in turn results in a different nonlinear refractive index. \Citet{Wang2013} uses a p-doped material with a doping concentration of \SI{1e15}{\per\cubic\centi\meter}, which corresponds to a surface resistivity of $ \approx\,\SI{13.5}{\ohm\centi\meter}$. \Citet{Lin2007} uses a similar p-doped crystal, but with a surface resistivity of \SI{20}{\ohm\centi\meter}, which corresponds to a dopant concentration of $ \approx\SI{6.7e14}{\per\cubic\centi\meter} $. Therefore, a lower dopant concentration in a p-doped crystal corresponds to a lower nonlinear refractive index.
	
	Those parameters can be combined in the so-called \emph{Nonlinear Figure of Merit} (NFOM). It usually is defined as the relative merit of the Kerr nonlinearity versus the two-photon nonlinearity by the ratio of both values, divided by the optical wavelength in vacuum, according to~\citet{Mizrahi1989}. When extending this ratio to multi-photon absorption, one obtains the NFoM in the following form~\cite{Wang2013}:

	\begin{equation}\label{equ:NFOM}
		\text{NFOM}=\left\{\begin{matrix}\frac{n_2}{\lambda\beta^{(2)}}&\forall\lambda\leq\SI{2200}{\nano\meter}\\
			\frac{n_2}{\lambda\beta^{(3)}I}&\forall\lambda>\SI{2200}{\nano\meter}\end{matrix}\right.,
	\end{equation}
	with $ \lambda $ the wavelength in vacuum, $ n_2(\lambda) $ the wavelength-dependent nonlinear refractive index, $ I $ the field intensity and $ \beta^{(K)} $ the coefficient for $ K $-photon absorption.
	
	As shown in \Cref{fig:nPA-Absorption} it has to be kept in mind that the three-photon absorption coefficient does not decrease to zero for wavelengths below \SI{2200}{\nano\meter}. It still can influence the laser light, provided a certain intensity is reached. This intensity can be determined by comparing the contribution of both two- and three-photon absorption and setting
	\begin{equation}\label{equ:ComparingBeta2Gamma}
		\beta^{(2)}\approx\beta^{(3)}I.
	\end{equation}
	For \SI{1550}{\nano\meter} this threshold is reached at $ I_\text{\SI{1550}{\nano\meter}}=\SI{2.8}{\peta\watt\per\square\meter}$, while for \SI{2150}{\nano\meter} the threshold can be calculated to be $ I_\text{\SI{2150}{\nano\meter}}=\SI{10}{\tera\watt\per\square\meter} $, and thus significantly lower. When no absorption is intended (such as in waveguides) the intensity typically stays below \SI{10}{\tera\watt\per\square\meter}~\cite{Gholami2011}, and therefore three-photon absorption can be neglected for shorter wavelengths. For those intensities \Cref{equ:NFOM} therefore stays valid.
	
	For processing the absorption, on the contrary, the localized absorption of laser pulses is the main aim, and therefore the intensities will be much higher at least close to the focal spot. At those intensities three-photon absorption will still be an important factor at wavelengths shorter than \SI{2200}{\nano\meter}, and therefore we propose a revised approach for the nonlinear figure of merit for large intensities:
	
	\begin{equation}\label{equ:RevisedNFOMForHighIntensity}
		\text{NFOM}_\text{rev}(\lambda, I_0)=\frac{n_2(\lambda)}{\lambda\left(\beta^{(2)} + \beta^{(3)}I_0\right)}.
	\end{equation}
	
	Similarly to three-photon absorption higher-order absorption processes will be present at shorter wavelengths, too. Therefore, it can be beneficial to verify if those processes have to be included in the revised nonlinear figure of merit, and potentially forming the following definition:
	
	\begin{equation}\label{equ:RevisedNFOMForVeryHighIntensity}
		\text{NFOM}_\text{revHigh}(\lambda, I_0)=\frac{n_2(\lambda)}{\lambda\sum_{K=2}^\infty\left(\beta^{(K)}I^{K-2}_0\right)}.
	\end{equation}
	
\begin{figure}[htp]
		\centering
		\includegraphics[width=.9\linewidth, height=.6\linewidth]{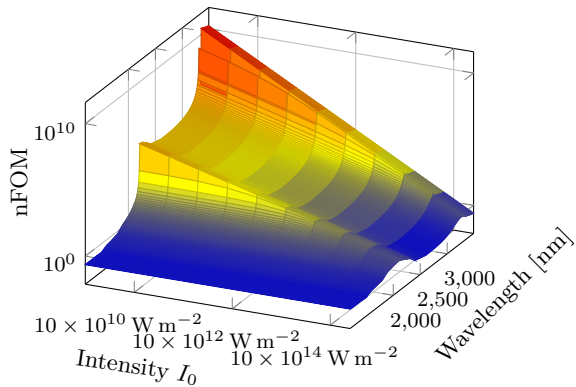}
		\caption{Nonlinear figure of merit (NFOM) as defined by \Cref{equ:RevisedNFOMForVeryHighIntensity}, drawn against different intensity- and wavelength values}
		\label{fig:3dNFOM}
	\end{figure}

	The calculation of the threshold intensity at which higher-order multi-photon coefficients will be of interest is complicated by sparse availability of measurement data for higher-order absorption coefficients in \ce{Si} over a large range of wavelengths. \Citet{Gai2013} provides several measurement points for the four-, five- and six-photon absorption coefficients, but only within a limited wavelength range (unlike the data given for the three-photon absorption coefficient). The lowest wavelength where the four-photon absorption coefficient has been measured is \SI{3250}{\nano\meter}~\cite{Gai2013}, with a value of \SI[separate-uncertainty=true]{1\pm0.25e-41}{\meter^5\per\watt\cubed}. Compared to the three-photon absorption coefficients at \SI{2150}{\nano\meter} and \SI{2550}{\nano\meter} a threshold intensity can be calculated (by following \Cref{equ:ComparingBeta2Gamma}) at $ I_0\geq\SI{2}{\peta\watt\per\square\meter} $. Together with the decrease of the four-photon absorption coefficient for shorter wavelengths (i.e. $ \lambda\leq\SI{3250}{\nano\meter} $) it therefore can be safely assumed that for intensities below \SI{2}{\peta\watt\per\square\meter} and above \SI{10}{\tera\watt\per\square\meter} \Cref{equ:RevisedNFOMForHighIntensity} can be used, and replaces \Cref{equ:NFOM} as the definition of the nonlinear figure of merit. Graphs based on \Cref{equ:NFOM,equ:RevisedNFOMForHighIntensity} showing the change of the nonlinear figure of merit for increasing intensities are given in \Cref{fig:3dNFOM}. Following the estimations made above the transition from \Cref{equ:NFOM} to \Cref{equ:RevisedNFOMForHighIntensity} is at $ I_0\geq\SI{1e13}{\watt\per\square\meter} $. The values for the nonlinear figure of merit depending on the intensity and wavelength are shown in \Cref{fig:3dNFOM}, with a cross-section at $ I_0=\SI{1e13}{\watt\per\square\meter} $ shown in \Cref{fig:NonFig1e13InText}. It has a clearly visible peak at $ \approx\SI{2200}{\nano\meter} $ due to the transition from two-photon absorption to three-photon absorption combined with the maximal Kerr-effect contribution.
	
	\begin{figure}[htp]
		\centering
		\includegraphics[width=.9\linewidth,height=.4\linewidth]{graphs/nfom_1e13_new.tikz}
		\caption{Nonlinear figure of merit for $ I_0=\SI{1e13}{\watt\per\square\meter} $, based on \Cref{fig:3dNFOM}}
		\label{fig:NonFig1e13InText}
	\end{figure}
	
	In \Cref{fig:3dNFOM} it can be seen that the three-photon absorption combined with the intensity gradually increases, which shifts the peak of the nonlinear figure of merit towards the shorter wavelengths (up to \SI{1950}{\nano\meter} for \SI{1}{\peta\watt\per\square\meter}) while simultaneously lowering the whole nonlinear figure of merit. This indicates the growing importance of higher-order absorption compared to lower-order absorption due to the scaling with the intensity, a relation which also is displayed in \Cref{fig:CurrentComparisonMPA}.
	
	\subsection{Numerical model and methods}
	\label{subsec:NumModel}
	To model the propagation of the pulse through Silicon we use the scalar unidirectional pulse propagation equation (UPPE)~\cite{Kolesik2002,Couairon2011}, written as 
	\begin{equation}\label{equ:UPPEScalar}
		\begin{split}
		\partial_z\hat{E}(\omega, z) = ik_z\hat{E}(\omega, z) &+ \frac{i\omega^2}{2\varepsilon_0c^2k_z}\hat{P}(\omega, z)\\
		&-\frac{\omega}{2\varepsilon_0c^2k_z}\hat{J}(\omega, z),
		\end{split}
	\end{equation}
	with $ \hat{E} $ the Fourier-transformed electric field of the pulse, $P$ the (non-)linear polarizations acting on the pulse and $J$ the (non-)linear current densities acting on the propagating pulse. The wave number is given by $\omega$ and the propagation constant $ k_z $ in a cylindrical coordinate system is defined as (based on~\citet{Couairon2015,Fedorov2016}) 
	\begin{equation}\label{equ:k_z}
		k_z = k_0 - \frac{k_\perp^2}{2k_0},
	\end{equation}
	to allow the calculation of tightly focusing optics with an NA $ \geq\num{0.2} $ while still using a scalar equation.
	
	The main advantage of the UPPE compared to the nonlinear Schrödinger equation (NLSE) for simulating the nonlinear propagation of a light pulse in Silicon is the inclusion of non-paraxial effects. This allows the numerical simulation of optics with a very large numerical aperture. 
	
	One has to note, that \Cref{equ:UPPEScalar,equ:k_z} correspond to the generalised Helmholtz equation written in a co-moving coordinate system $(z = z,\,\,t = T - z/u)$ ($t$ is a local time, $T$ is a time in the laboratory coordinate system, $u$ is a group velocity of a pulse, and $r$ is a radial coordinate within the axial symmetry approximation)~\cite{PhysRevLett.76.4356}:
	
	\begin{equation}\label{equ:Helmholz}
		\left[ {{\partial _z} + \frac{i}{{2{k_0}}}\left( {\partial_z^2 + {\nabla^{2}_\perp }} \right)} \right]E\left( {t,z,r} \right) = \Xi \left\langle {E\left( {t,z,r} \right)} \right\rangle.
	\end{equation}
	
	The transverse Laplacian in \Cref{equ:Helmholz} reads as $ \nabla^{2}_\perp\to\partial^2/\partial r^2 + (1/r)\partial/\partial r $, where $r$ is a radial coordinate. 
	
	The contribution $\Xi \left\langle {E\left( {t,z,r} \right)} \right\rangle$ to the nonlinear polarization corresponds to that represented by the terms in \cref{equ:UPPEScalar}. That are defined by the Kerr-effect, described in \cref{subsec:PhysNonlinParams}, while the nonlinear multi-photon absorption (two- and three-photon absorption) and the plasma absorption and -defocusing are part of the nonlinear currents.
	
	The UPPE-equation (c.f. \Cref{equ:UPPEScalar,equ:k_z}) is a very effective tool for simulations of field dynamics outside the paraxial approximation. Using the split-step Fourier approach could promise an acceleration of computation. However, the interweaving of spatial and space degrees of freedom requires the evaluation of big-size matrices, which is a memory-intensive task. The generalized Helmholtz-equation (\Cref{equ:Helmholz}) is evaluated based on the finite element method (FEM) in the time domain. The advantage of this technique is its applicability to complicated geometrical structures, the possibility of using a broad class of very effective numerical solvers, and the adaption of the simulation mesh which can accelerate calculations substantially. Both methods are based on the same equation set and are complementary. Their preferable use depends on the conditions of strongly unsteady-state dynamics~\cite{chang1999difference,Bertolani}. Nevertheless, one has to note that FEM allows adopting the mesh to a concrete field configuration changing with propagation and, thereby, accelerating the calculations substantially. In this paper we used both methods to make use of the best complementarity of both approaches, and maximum efficiency of numerical calculations. 
	
	Finally, the generation of free carriers is describes by using (from~\citet{Couairon2011})
	\begin{equation}\label{equ:GenerationFreeCarriers}
		\frac{\partial\varrho}{\partial t}=W_{\text{ofi}}(I)\left(\varrho_{nt} - \varrho\right) + W_{\text{ava}}(I)\varrho.
	\end{equation}
	The rate equation contains the free carriers generated due to nonlinear absorption, defined as
	\begin{equation}\label{equ:Wofi}
		W_{\text{ofi}}=\sigma_KI^K,
	\end{equation}
	with $K$ the number of involved photons, and the free carriers generated due to avalanche ionization, defined as
	\begin{equation}\label{equ:Wava}
		W_{\text{ava}}=\sigma(\omega_0)\frac{I}{E_g}.
	\end{equation}
	Here, the inverse Bremsstrahlung coefficient $ \sigma(\omega_0) $ is defined as
	\begin{equation}\label{equ:InvBremsCoef}
		\sigma(\omega_0)=\frac{\omega_0}{n(\omega)c\frac{\varepsilon_0m_e\omega_0^2}{q_e^2}}\frac{\omega_0\tau_c\left(1+i\omega\tau_c\right)}{1+\omega^2\tau_c^2}.
	\end{equation}
	The intensity $I$ is defined as $I=\frac{cn\varepsilon_0}{2}\vert E\vert^2$, and $E_g$ is the band gap energy. $ \tau_c $ in \Cref{equ:InvBremsCoef} is the electron collision time. Finally, $ \sigma_K $ in \Cref{equ:Wofi} is given as 
	\begin{equation}\label{equ:SigmaK}
		\sigma_K=\frac{\beta^{(K)}}{K\hbar\omega_0\varrho_{nt}}\text{ \cite{Couairon2011},}
	\end{equation}
	with $ \varrho_{nt} $ the neutral free carrier density.
	
	We decided to neglect the recombination rate of the generated free carriers due to their long lifetime, up to $ \approx\SI{10}{\nano\second} $~\cite{Kononenko2012}, which is significantly longer than the pulse duration.
	
	\Cref{equ:UPPEScalar} was implemented while using functions from the libraries deal.II~\cite{dealII93, dealii2019design} and PETSc~\cite{petsc-efficient,petsc-user-ref}. The computations were run on the local HPC cluster IDUN/EPIC~\cite{sjalander+:2019epic} and on resources provided by UNINETT Sigma2 -- the National Infrastructure for High Performance Computing and Data Storage in Norway.
	
	\section{Numerical simulations}
	\label{sec:NumSims}
	Initial investigations of the wavelength dependence of the pulse propagation in silicon was already done before, as for example by~\citet{Richter2018,Richter2020,Richter2021}, but those investigations were limited by the used method for numerical simulations. \Citet{Richter2018,Richter2020} used the NLSE for simulating the pulse propagation, which limits the maximum focusing angle which can be used. On the other hand~\citet{Das2020} reported experimentally for $ \lambda=\SI{1550}{\nano\meter} $ that an NA value of $ \geq\num{0.85} $ has to be used for pulse durations of $ \approx\SI{5}{\pico\second} $, which limits the applicability of the NLSE for those pulse durations. Nevertheless, at least for lower focusing angles it could be shown that between \SIrange[range-units=single, range-phrase=\text{ to }]{2000}{2200}{\nano\meter} the efficiency of the energy delivery to the focal spot was increased compared to longer and shorter wavelengths.
	
	To investigate and correlate the experimentally observed behavior with numerical simulations we chose to simulate a pulse at four different wavelengths, $ \SI{1550}{\nano\meter} $, $ \SI{1950}{\nano\meter} $, $ \SI{2150}{\nano\meter} $ and $ \SI{2350}{\nano\meter} $ with the pulse energies ranging from \SI{0.001}{\micro\joule} to \SI{1}{\micro\joule}. A full overview over all parameters is given in \Cref{tab:OverviewPulseLensParameter}. Here, the maximum pulse energy for an NA of \num{0.25} is \SI{500}{\nano\joule}, while the maximum pulse energy for an NA of \num{0.85} is \SI{5000}{\nano\joule}.
	
	\begin{table}[htpb]
		\centering
		\caption{Overview over the used pulse- and lens parameters in the numerical simulations}
		\label{tab:OverviewPulseLensParameter}
		\begin{tabular}{lll}
			\toprule
			Parameter&Value&Unit\\
			\midrule
			Pulse duration $ t_0 $&\num{100}, \num{500}, \num{1000}, \num{5000}&\si{\femto\second}\\
			Pulse wavelength $ \lambda_0 $&\leftspecialcell{\num{1550}, \num{1950}, \num{2000},\\ \num{2150}, \num{2550}, \num{3250}}&\si{\nano\meter}\\
			Pulse energy $ E_0 $&\leftspecialcell{\num{1}, \num{10}, \num{62.5}, \num{100},\\\num{500}, \num{1000}, \num{5000}}&\si{\nano\joule}\\
			\leftspecialcell{Numerical aperture\\
				at \SI{1550}{\nano\meter}}&\num{0.1}, \num{0.25}, \num{0.85}&\\
			\leftspecialcell{Beam radius\\at \SI{1550}{\nano\meter}}&1&\si{\milli\meter}\\
			Focus depth&\num{100}, \num{250}, \num{500}, \num{700}&\si{\micro\meter}\\
			\bottomrule
		\end{tabular}
	\end{table}
	
	For all calculations with the UPPE the initial beam diameter was set to $d_0=\SI{2}{\milli\meter}$ for a wavelength of \SI{1550}{\nano\meter}. To achieve a similar fluence in the focal spot for linear propagation, the beam radius $ r_\lambda $ for other wavelengths was adjusted using
	\begin{equation}\label{equ:BeamDiamAdjustment}
		r_\lambda=\frac{d_0}{2}\cdot\frac{\lambda}{\SI{1550}{\nano\meter}},
	\end{equation}
	unless explicitly noted otherwise.
	
	While this will result in a similar fluence level at the focal spot independently of the wavelength, it will result in changed focusing angles. This in turn will influence the impact of the nonlinear effects onto the propagation of the pulse. A smaller NA value will result in a higher intensity closer to the surface of the material, and thereby triggering self-focusing and nonlinear absorption earlier compared to larger focusing angles. The used NA-values corresponding to each wavelength are listed in \Cref{tab:AdjustedNAValues}.
	
	\begin{table}[htp]
		\centering
		\caption{NA-values used for the wavelengths displayed in \Cref{tab:OverviewPulseLensParameter} when using \Cref{equ:UPPEScalar} as pulse propagation equation, following \Cref{equ:BeamDiamAdjustment}, with an initial beam diameter of $ \SI{2}{\milli\meter} $}
		\label{tab:AdjustedNAValues}
		\begin{tabular}{c|c|c}
			\toprule
			Wavelength&Target NA: \num{0.25}&Target NA: \num{0.85}\\
			\midrule
			\SI{1550}{\nano\meter}&0.25&0.85\\
			\SI{1950}{\nano\meter}&0.31&1.07\\
			\SI{2150}{\nano\meter}&0.35&1.18\\
			\SI{2550}{\nano\meter}&0.41&1.40\\
			\SI{3250}{\nano\meter}&0.52&1.78\\
			\bottomrule
		\end{tabular}
	\end{table}
	
	The numerical apertures used for the HHE are listed in \Cref{tab:NAForHHE}, in comparison.
	
	\begin{table}[htp]
		\centering
		\caption{NA-values used for the wavelengths displayed in \Cref{tab:OverviewPulseLensParameter} when using \Cref{equ:Helmholz} as pulse propagation equation}
		\label{tab:NAForHHE}
		\begin{tabular}{c|c|c}
			\toprule
			Wavelength&Numerical aperture&Figure\\
			\midrule
			\SI{1550}{\nano\meter}&\num{0.25}&\Cref{fig:FunnyLightRings,fig:MultiFigWithPlasmaInfluence}\\
			\SI{2000}{\nano\meter}&\num{0.1}&\Cref{fig:ComparingParaxialWithNonparaxial}\\
			\SI{3250}{\nano\meter}&\num{0.25}, \num{0.85}&\Cref{fig:longwavelengthFigure}\\
			\bottomrule
		\end{tabular}
	\end{table}
	\subsection{Impact of the intensity on the absorption}
	\label{subsec:ImpactIntensityAbsorption}
\begin{figure}[htp]
		\centering
		\includegraphics[width=\linewidth,height=.7\linewidth]{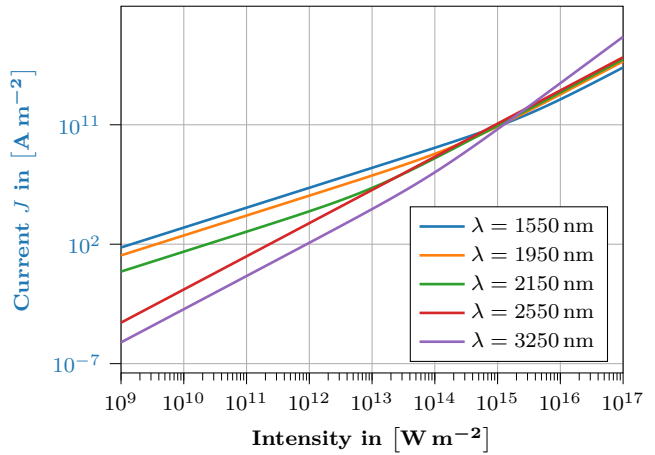}
		\caption{Calculation of the current induced by multi-photon absorption, following \Cref{equ:ContributionMPAtoCurrent}, by using the wavelengths and the corresponding nonlinear absorption coefficients and refractive indices displayed in \Cref{tab:PhysParamsExample}. The current for \SI{1550}{\nano\meter} is drawn in blue, and the current for \SI{1950}{\nano\meter} is drawn in orange. For \SI{2150}{\nano\meter} the current is shown in green, for \SI{2550}{\nano\meter} it is shown in red, and for \SI{3250}{\nano\meter} in purple.}
		\label{fig:CurrentComparisonMPA}
	\end{figure}

	The multi-photon absorption coefficients of order $ K $ shown in \Cref{fig:nPA-Absorption} contribute to the current density $ J $ as defined in \Cref{equ:UPPEScalar} via
	\begin{equation}\label{equ:ContributionMPAtoCurrent}
		J_K=\varepsilon_0cn_0\beta^{(K)}I^{K-1}E\text{, \cite{Couairon2011},}
	\end{equation}
	with $ n_0 $ the refractive index, $ \beta^{(K)} $ the absorption coefficient of $ K $-photon absorption and $ I $ the intensity. While the absorption due to free carriers does scale with the wavelength, but not with the intensity, and therefore is not included in this discussion. Combining \Cref{equ:ContributionMPAtoCurrent} with \Cref{tab:PhysParamsExample} the generated current can be calculated, which is displayed in \Cref{fig:CurrentComparisonMPA}, and will be shortly elaborated in \Cref{subsec:ImpactIntensityAbsorption}.
	
	As displayed in this figure the total absorption due to multi-photon absorption scales with the intensity. Depending on the order of the multi-photon absorption this scaling factor is either purely $ I^2 $, $ I^3 $ or a combination of both. A higher scaling factor can lead to a higher absorption current, even though the absorption coefficient itself is lower. This effect can be seen especially for \SI{2550}{\nano\meter}. While the generated current for an intensity of \SI{1e9}{\watt\per\meter\squared} is five orders of magnitude lower than the current generated by \SI{2150}{\nano\meter}, it is equal for an intensity of $\approx\SI{3e13}{\watt\per\meter\squared}$, and higher than that for increasing intensities. These intensities are roughly comparable to the intensities reached by a pulse with an energy of \SI{1}{\nano\joule} and a duration of \SI{100}{\femto\second} at a wavelength of \SI{2150}{\nano\meter}, focused with a lens with $ f=\SI{4}{\milli\meter} $ and a beam diameter of \SI{1}{\milli\meter}. This gives a focal spot area of $\approx\SI{90}{\micro\meter\squared}$, and with a peak power (for a Gaussian pulse) of \SI{8.8}{\kilo\watt} the peak intensity is $\approx\SI{9.4e13}{\watt\per\meter\squared}$, i.e. an intensity which can be easily achieved.
	
	Similarly, the transition intensity for the shorter wavelengths can be taken from \Cref{fig:CurrentComparisonMPA}, and can be approximated to be $\approx\SI{1e15}{\watt\per\square\meter} $, corresponding to a pulse with $ \approx\SI{10}{\nano\joule} $ and the parameters used above. Nevertheless, one has to keep in mind that the intensity at the focal spot was taken into account here. When focusing deep into the material the affected area increases, which in turn decreases the intensity. Setting the focal depth to \SI{500}{\micro\meter} in \ce{Si}, corresponding to $ \approx\SI{144}{\micro\meter} $ in air, will increase the affected area by a factor of $\approx\num{11} $, and reduce the intensity accordingly. 
	Decreasing the focal length of the lens (and thereby increasing the numerical aperture) will affect the intensity. Going from an NA of \num{0.125} as used above to an NA of \num{0.85} (which is typically used for processing experiments~\cite{Richter2020,Das2020}) by decreasing $ f $ from \SI{4}{\milli\meter} to \SI{0.58}{\milli\meter} will increase the affected area by $ \approx\num{130} $ and lower the intensity accordingly. This is especially important for longer wavelengths, after the total absorption for higher-order multi-photon absorption processes scales faster with the intensity.
	\subsection{Influence of the non-paraxiality on the pulse shape}
	\label{subsec:InfluenceNonParaxiality}
	To demonstrate the importance of using methods which do not rely on the paraxial approximation for propagating the laser pulse a pulse was simulated using \Cref{equ:Helmholz}, once while including the paraxial approximation, and once without, while keeping the pulse parameters constant. The pulse structure at the same position within silicon is shown in \Cref{subfig:ComparingApproximationsParaxial} for the case of included paraxial approximation, and in \Cref{subfig:ComparionApproximationsNonParaxial} for the case without approximation. 
	
	\begin{figure}[htp]
		\centering
		\begin{subfigure}[t]{.9\linewidth}
			\centering
			\includegraphics[width=\linewidth, height=.4\linewidth]{images/Images_Vladimir/fig_7_paraxial.tikz}
			\caption{Paraxial approximation}
			\label{subfig:ComparingApproximationsParaxial}
		\end{subfigure}\hfill%
		\begin{subfigure}[t]{.9\linewidth}
			\centering
			\includegraphics[width=\linewidth, height=.4\linewidth]{images/Images_Vladimir/fig_7_nonparaxial.tikz}
			\caption{Non-paraxial approximation}
			\label{subfig:ComparionApproximationsNonParaxial}
		\end{subfigure}%
		\caption{Contour plots of intensity demonstrating the time profile asymmetry due to plasma generation. The pulse parameters are (following \Cref{equ:InitialConditionsHelmholtzDescription}) $ P_0=\SI{0.5}{\mega\watt} $, $ f=\SI{700}{\micro\meter} $, NA = \num{0.1}, $ w_0 = \SI{6.4}{\micro\meter} $, $ \lambda=\SI{2000}{\nano\meter} $, and $ t_0=\SI{250}{\femto\second} $. The displayed field profiles are taken in a depth of \SI{670}{\micro\meter} below the surface.}
		\label{fig:ComparingParaxialWithNonparaxial}
	\end{figure}
	
	In comparison a clear difference is visible. While the paraxial approach results in an elongated structure, the non-paraxial approach shows a clear intensity dip in the center of the pulse. Despite a relatively small NA~\Cref{fig:ComparingParaxialWithNonparaxial}, the non-paraxial approximation addresses more precisely the nonlinear factors causing the spatial-time reshaping of a beam.
	
	\subsection{Spatio-temporal structure and impact of spherical aberrations}
	\label{subsec:SpatiotemporalStructure}
	While propagating through the material the pulse changes its shape both in space and in time. This is visualized in \Cref{subfig:z1473,subfig:z1578,subfig:z1894,subfig:z2000}. These figures demonstrate the spatio-temporal evolution of a light pulse with $ t_0=\SI{100}{\femto\second} $, $ E_0=\SI{100}{\nano\joule} $ and a wavelength of $ \lambda=\SI{1550}{\nano\meter} $, focused into \ce{Si} with a lens with $ \text{NA}=\num{0.25} $. The focal spot was placed at \SI{500}{\micro\meter} in air, corresponding to \SI{1736}{\micro\meter} in \ce{Si} due to the increased refractive index.
	
	Simulations based on both \Cref{equ:UPPEScalar,equ:Helmholz} demonstrate the appearance of spherical aberrations (c.f. \Cref{subfig:LightRingZ160}). A spatial substructure appears in the form of \definedterm{light rings}, which evolves to $ r=\SI{0}{\micro\meter} $ and squeezes with a distance due to self-focusing (c.f. \Cref{subfig:LightRingZ360}).
	
	\begin{figure}[htp]
		\centering
		\begin{subfigure}[t]{.95\linewidth}
			\centering
			\includegraphics[width=\linewidth, height=.4\linewidth]{images/Images_Vladimir/fig_6a_cut.tikz}
			\caption{z=\SI{160}{\micro\meter}}
			\label{subfig:LightRingZ160}
		\end{subfigure}\hfill%
		\begin{subfigure}[t]{.95\linewidth}
			\centering
			\includegraphics[width=\linewidth, height=.4\linewidth]{images/Images_Vladimir/fig_6b_cut.tikz}
			\caption{z=\SI{360}{\micro\meter}}
			\label{subfig:LightRingZ360}
		\end{subfigure}%
		\caption{Contour plots of \definedterm{light rings} in the intensity (spherical aberrations) in dependence on the propagation distance inside the medium. The pulse parameters are (following \Cref{equ:InitialConditionsHelmholtzDescription}) $ P_0=\SI{10}{\mega\watt} $, $ f=\SI{500}{\micro\meter} $, NA = \num{0.25}, $ w_0=\SI{2}{\micro\meter} $, $ \lambda=\SI{1550}{\nano\meter} $, and $ t_0=\SI{100}{\femto\second} $}
		\label{fig:FunnyLightRings}
	\end{figure}
	
	\begin{figure}[!htp]
		\centering
		\begin{subfigure}[t]{.8\linewidth}
			\centering
			\includegraphics[width=\linewidth]{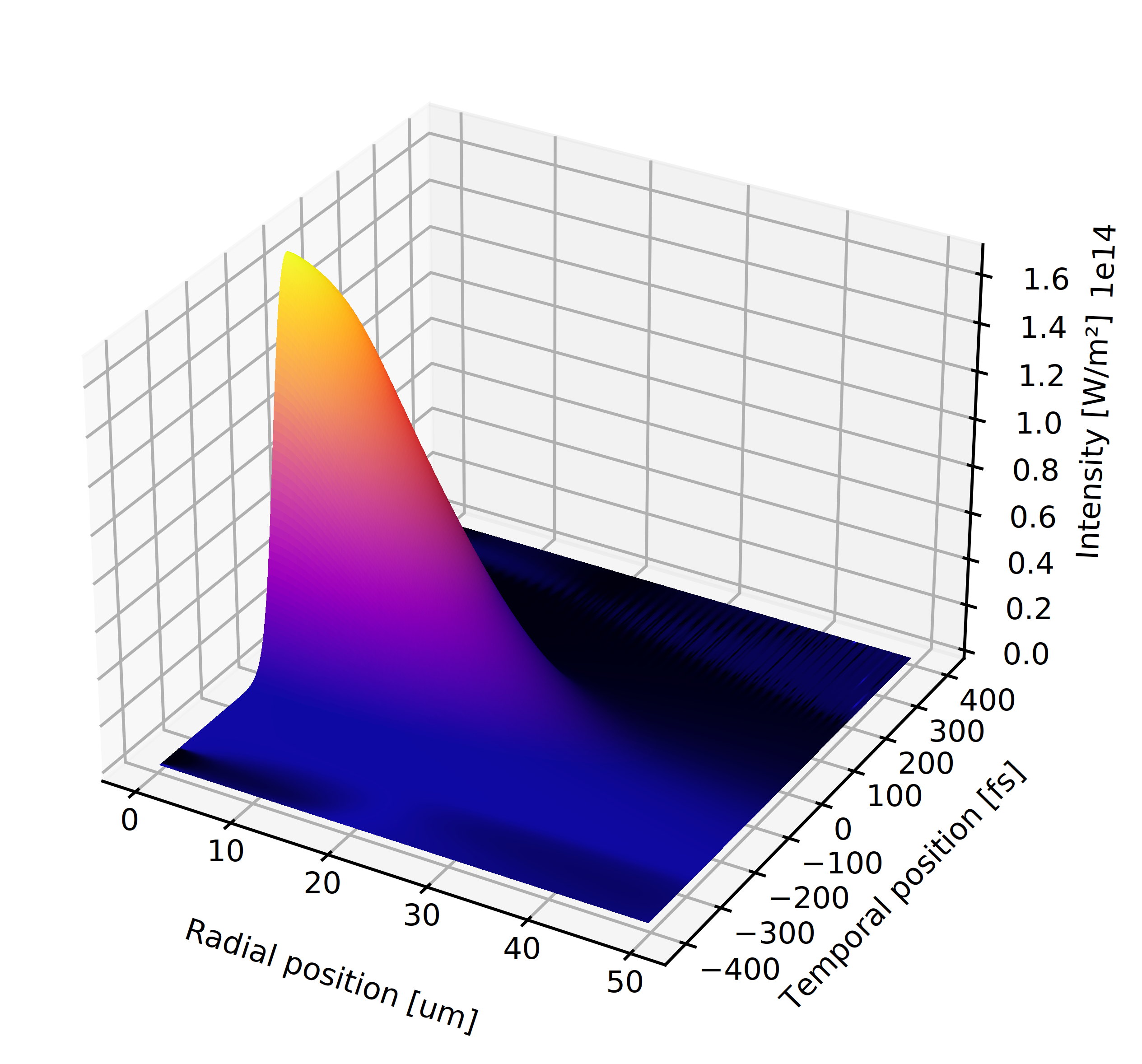}
			\caption{z = $ \SI{1473}{\micro\meter} $ in \ce{Si}, \SI{423}{\micro\meter} in air}
			\label{subfig:z1473}
		\end{subfigure}%
	
		\begin{subfigure}[t]{.8\linewidth}
			\centering
			\includegraphics[width=\linewidth]{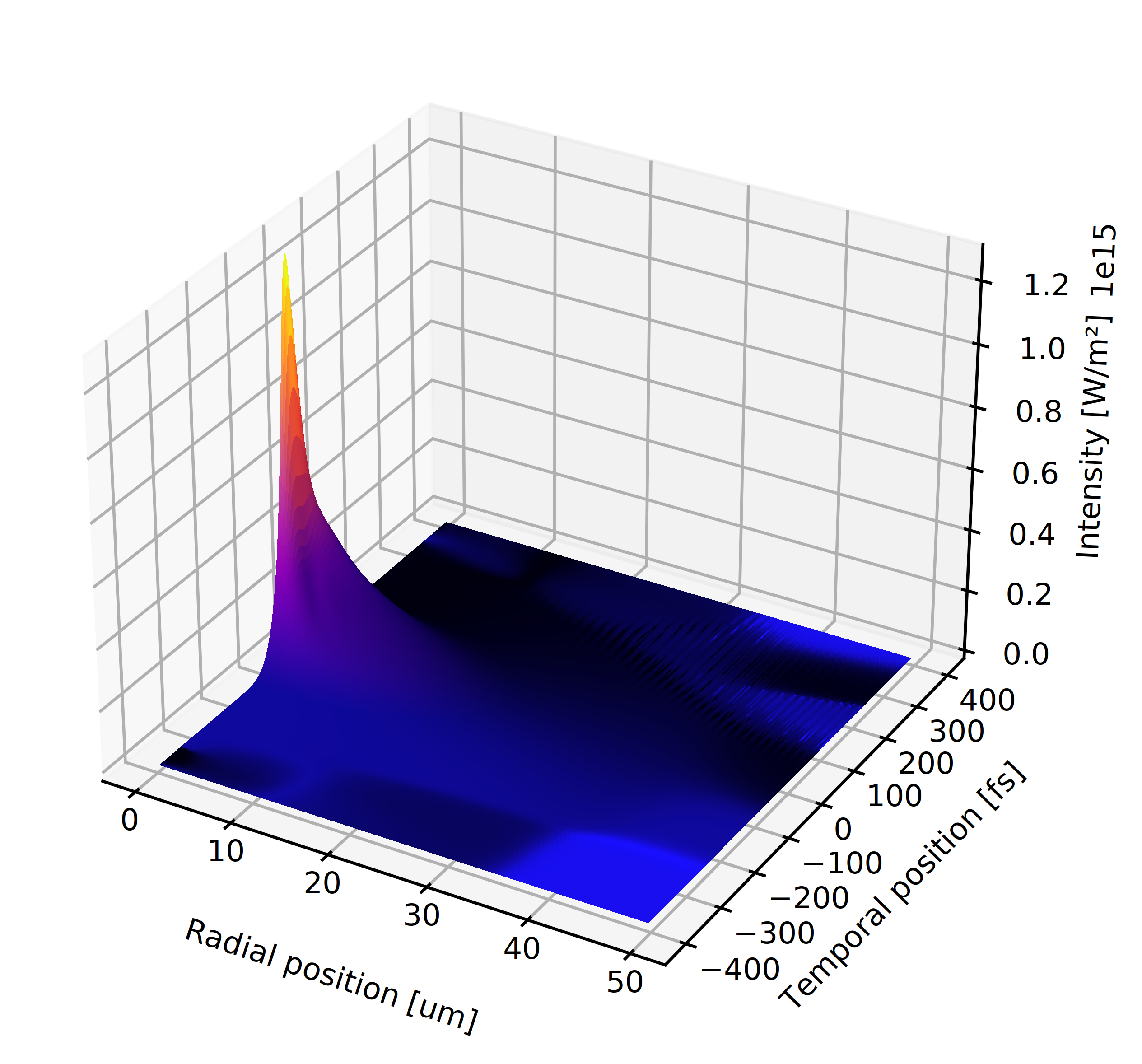}
			\caption{z = $ \SI{1578}{\micro\meter} $ in \ce{Si}, \SI{454}{\micro\meter} in air}
			\label{subfig:z1578}
		\end{subfigure}%
		
		\begin{subfigure}[t]{.8\linewidth}
			\centering
			\includegraphics[width=\linewidth]{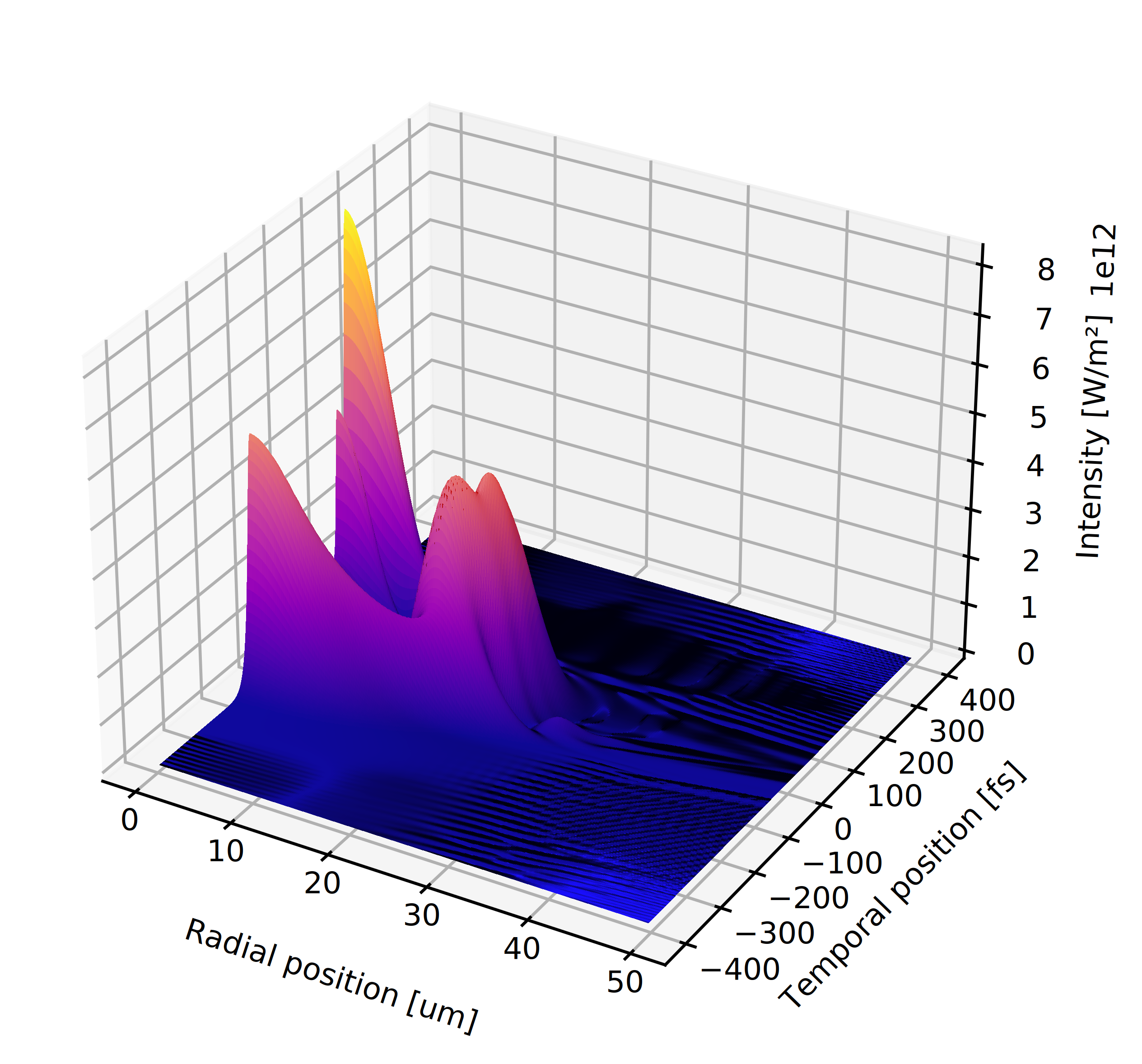}
			\caption{z = $ \SI{1894}{\micro\meter} $ in \ce{Si}, \SI{545}{\micro\meter} in air}
			\label{subfig:z1894}
		\end{subfigure}%
	\end{figure}

	\begin{figure}[!htp]
		\ContinuedFloat
		\centering
		\begin{subfigure}[t]{.8\linewidth}
			\centering
			\includegraphics[width=\linewidth]{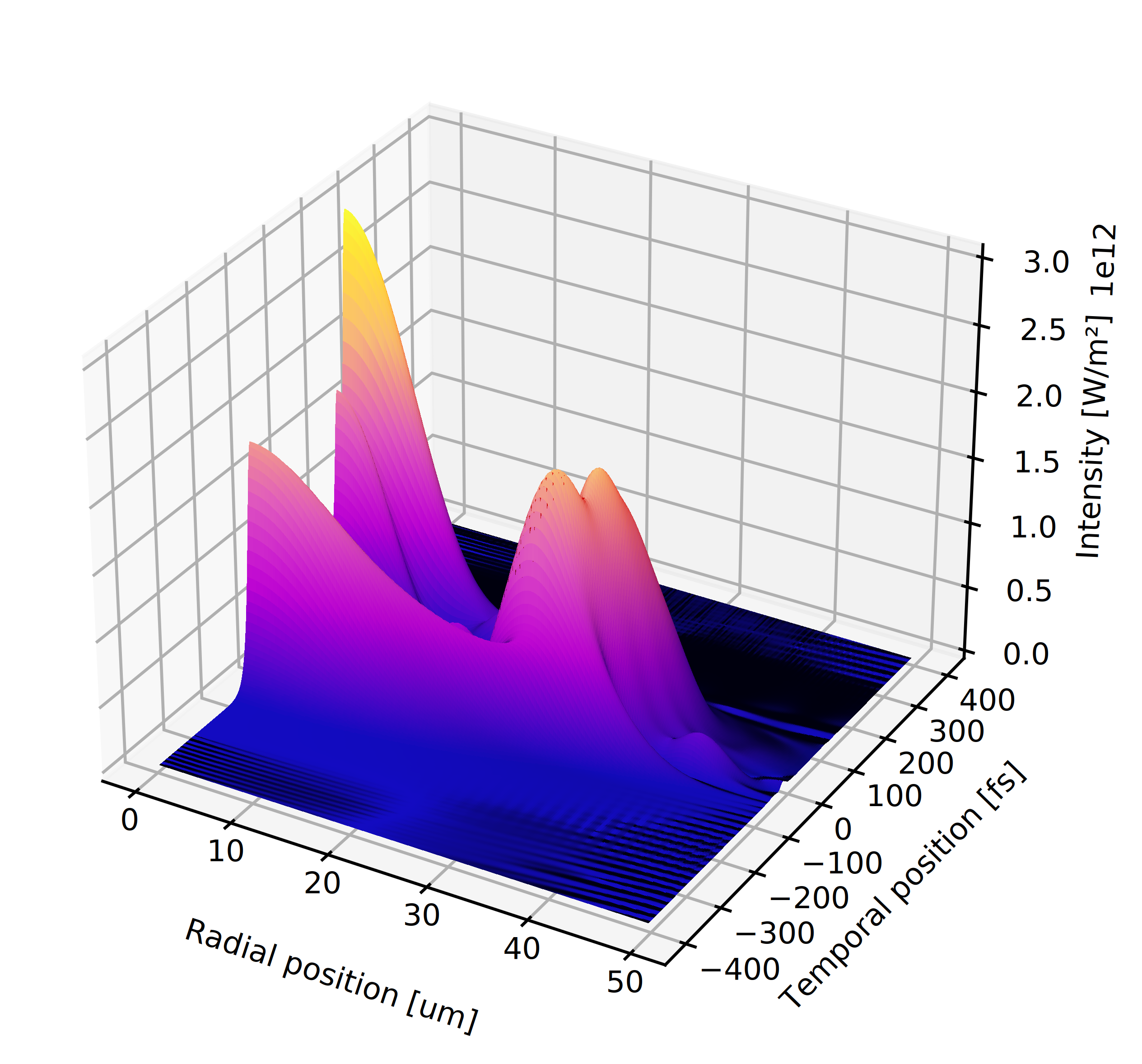}
			\caption{z = $ \SI{2000}{\micro\meter} $ in \ce{Si}, \SI{575}{\micro\meter} in air}
			\label{subfig:z2000}
		\end{subfigure}
		\caption{Spatio-temporal intensity field distribution of a pulse with $ \lambda=\SI{1550}{\nano\meter} $, $ E_0=\SI{100}{\nano\joule} $ and a duration of $ \SI{100}{\femto\second} $. The pulse is focused using a lens with $ f=\SI{500}{\micro\meter} $ and a numerical aperture of \num{0.25}. \Cref{subfig:z1578} shows the pulse structure shortly before the focal spot at $ z=\SI{500}{\micro\meter} $, \Cref{subfig:z1894} shortly after the focal spot, and \Cref{subfig:z2000} far away from the focal spot. The effect of self-focusing is clearly visible from \Cref{subfig:z1473} to \Cref{subfig:z1578}, followed by nonlinear absorption and plasma-induced defocusing after the focal spot.}
		\label{fig:SpatioTemporalShape}
	\end{figure}								
	Several points can be noted about the pulse behavior already before the focal spot. Going from \SI{423}{\micro\meter} (\Cref{subfig:z1473}) to \SI{454}{\micro\meter} (\Cref{subfig:z1578}) not only the pulse in total is focused towards the focal spot, but also the effect of self-focusing is visible. The innermost part of the pulse with the highest intensity is focused more compared to the rest of the pulse, creating an asymmetric shape along the spatial axis. Nevertheless, the pulse stays symmetric along the time axis.
	
	This effect can be considered as analogue to the Kerr-beam self-cleaning in multi-mode fibers, as demonstrated by~\citet{Krupa2019}. The calculations demonstrate that the NA growth enhances the tendency for generating multiple rings. Still, this effect degrades quickly at the wavelength close to the maximum of NFOM and longer wavelengths.
	
	Shortly before the focusing point, though, the self-focusing effect together with the generated plasma leads to a deformation of the beam not only in space, but also in time. As also shown below in \Cref{subfig:SliceIntensity025100100Comp,subfig:SliceIntensity025100250Comp,subfig:SliceIntensity085100100Comp,subfig:SliceIntensity085100250Comp}, the leading part of the pulse generates a large amount of free carriers, which lead to diffraction and multi-photon absorption. Due to the multi-photon absorption scaling nonlinearly with the intensity this affects the central part of the pulse dis-proportionally more compared to the leading and trailing edge. Generated free carriers also lead to absorption, but only linearly, and therefore can be neglected for high intensities. Still, for a sufficient density the absorption due to free carriers is visible, and can be seen for the trailing part of the pulse in \Cref{subfig:z1894,subfig:z2000}. For this part the intensity is reduced by $ \approx\SI{50}{\percent} $ compared to the leading edge.
	
	\Cref{fig:ComparingParaxialWithNonparaxial}, based on \Cref{equ:Helmholz}, demonstrates the effect of time asymmetry in the vicinity of the first maximum of the NFOM (c.f. \Cref{fig:3dNFOM,fig:NonFig1e13InText}) at $ \lambda=\SI{2000}{\nano\meter} $. This effect is connected with the plasma generated by the pulse front that initiates the pulse \actualquote{explosion} (c.f. \Cref{fig:MultiFigWithPlasmaInfluence}).
	
	\begin{figure}[!htb]
		\centering
		\begin{subfigure}[t]{\linewidth}
			\centering
			\includegraphics[width=.55\linewidth]{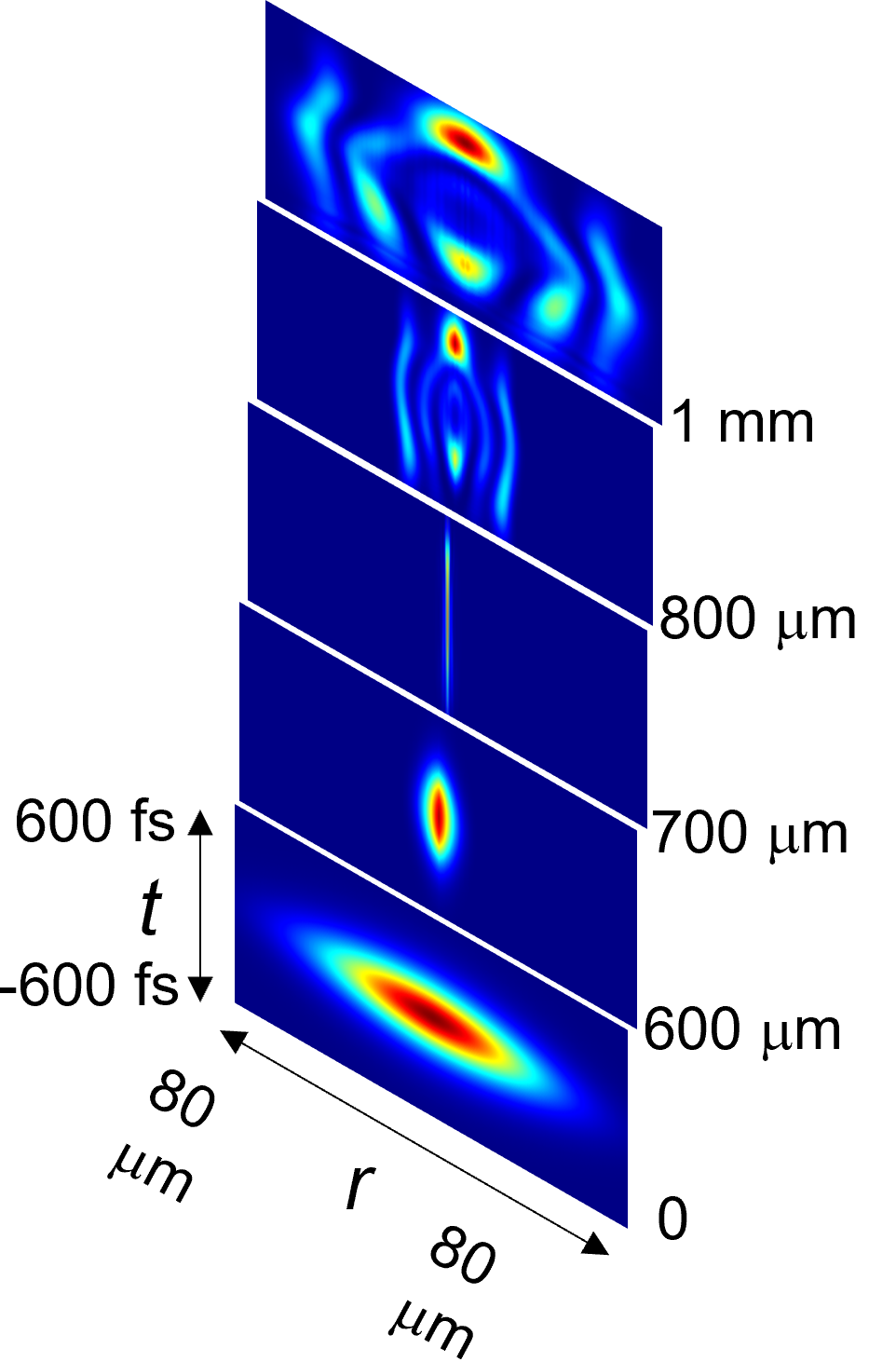}
			\caption{Temporal field distribution of the pulse at different positions inside the material.}
			\label{fig:MultiFigField}
		\end{subfigure}
	
		\begin{subfigure}[t]{\linewidth}
			\centering
			\includegraphics[width=.75\linewidth]{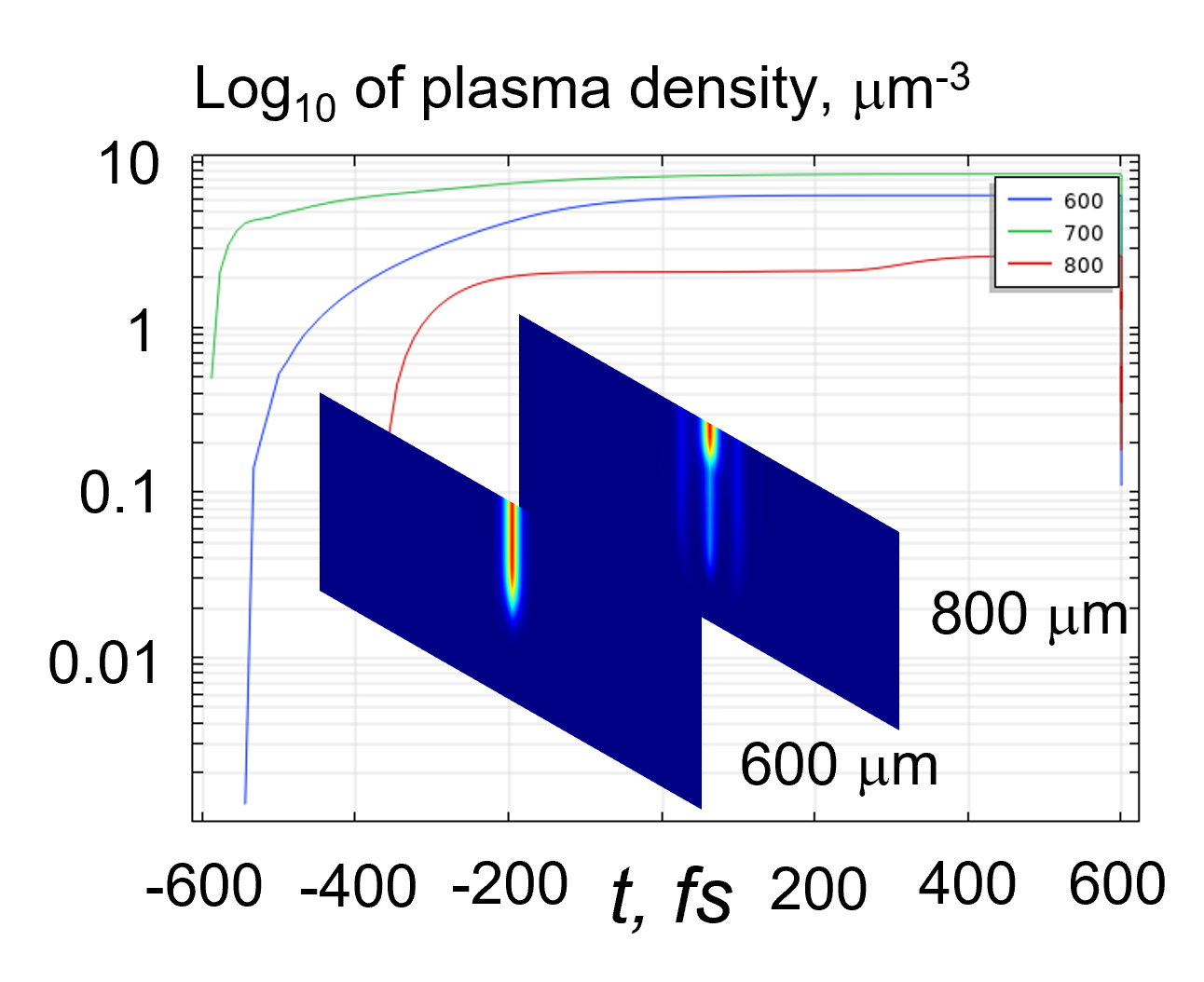}
			\caption{Contour plots and time dependence of plasma density at the beam axis, corresponding to \Cref{fig:MultiFigField}.}
			\label{fig:MultiFigFieldDensity}
		\end{subfigure}
	\setlength{\belowcaptionskip}{-10pt}
	\end{figure}

	\begin{figure}[!htb]
		\ContinuedFloat
		\centering
		\begin{subfigure}[t]{\linewidth}
			\centering
			\includegraphics[width=.75\linewidth]{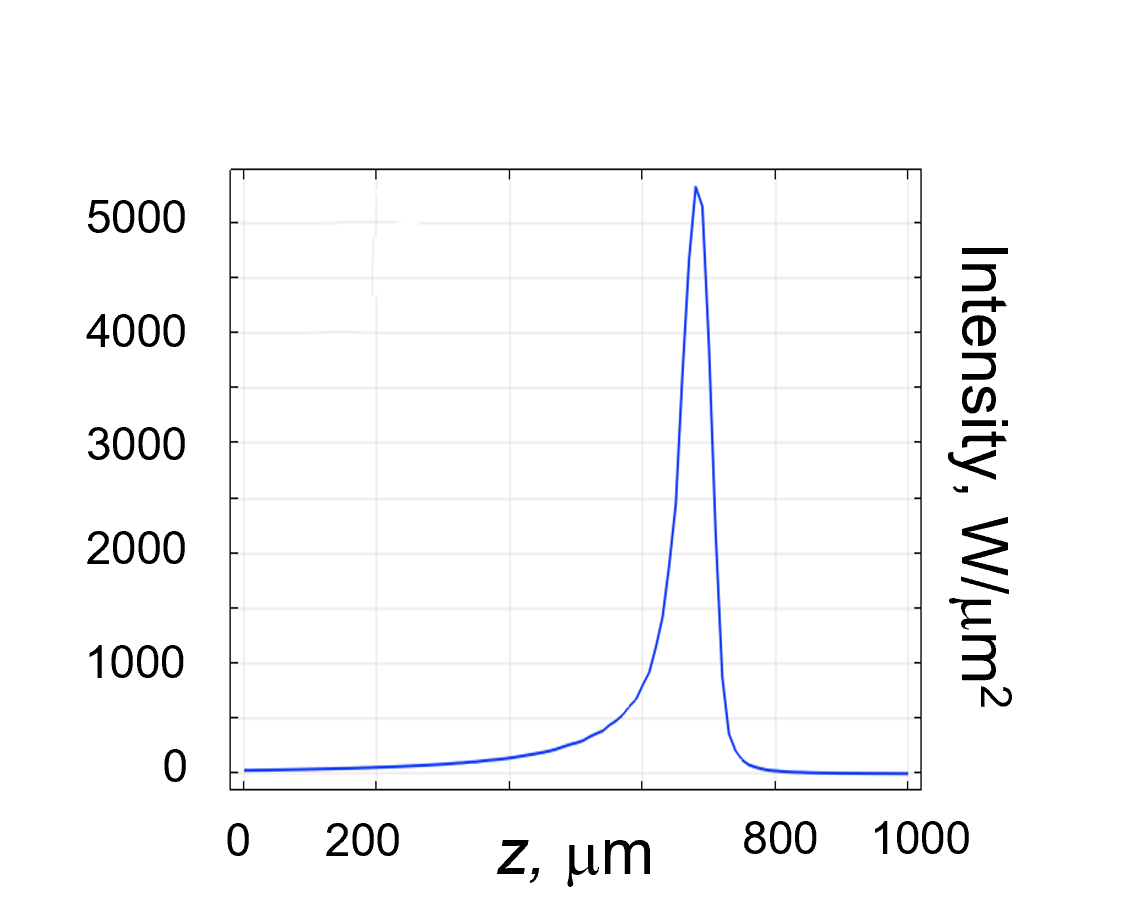}
			\caption{Intensity evolution.}
			\label{fig:MultiFigIntensity}
		\end{subfigure}%
		
		\begin{subfigure}[t]{\linewidth}
			\centering
			\includegraphics[width=.75\linewidth]{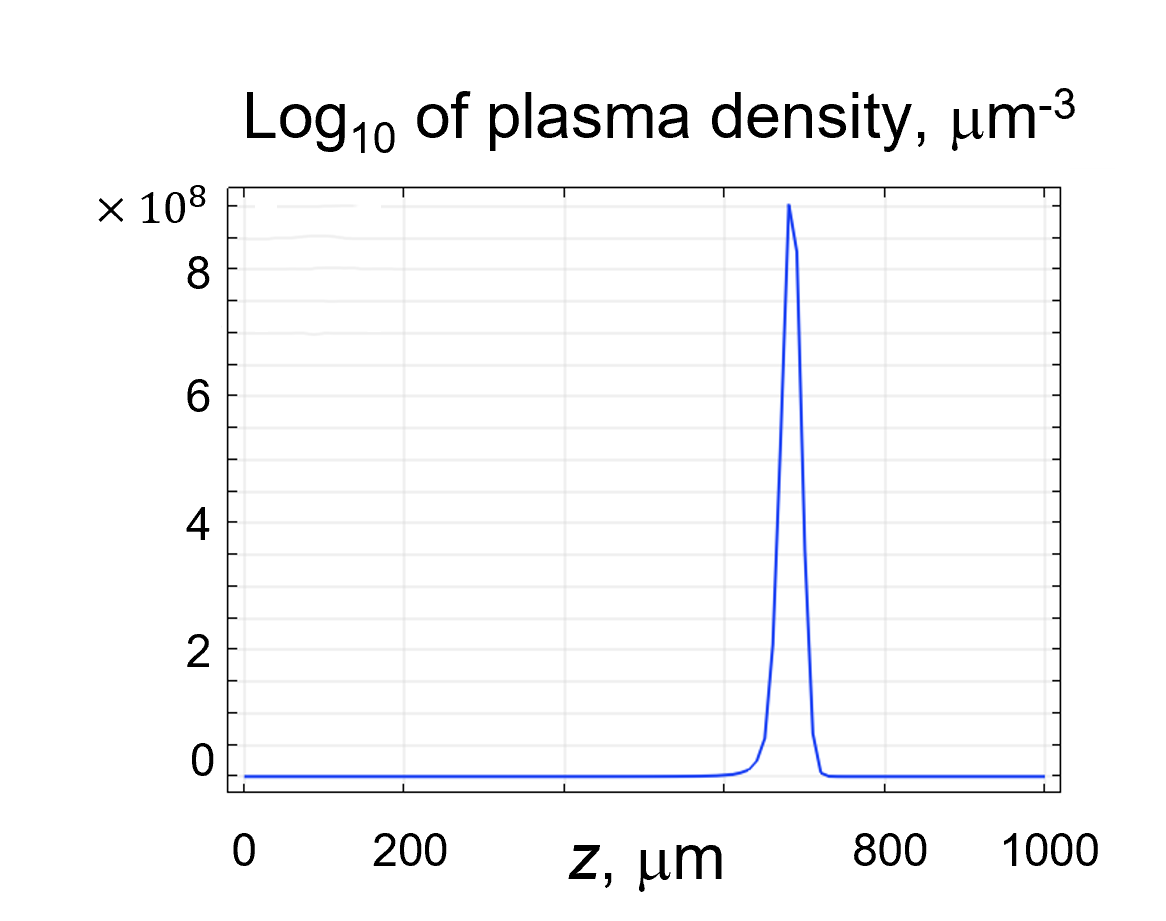}
			\caption{Density evolution.}
			\label{fig:MultiFigDensity}
		\end{subfigure}
		\setlength{\belowcaptionskip}{-10pt}
		\caption{Contour plots of the intensity evolution, demonstrating the time profile asymmetry and pulse explosion (\Cref{fig:MultiFigField}) due to plasma generation (corresponding contour plots and time dependence of plasma density at the beam axis are shown in \Cref{fig:MultiFigFieldDensity}). The intensity and the plasma density evolution are illustrated in \Cref{fig:MultiFigIntensity,fig:MultiFigDensity}, respectively. The parameters are the same as in \Cref{fig:FunnyLightRings}.}
		\label{fig:MultiFigWithPlasmaInfluence}
	\end{figure}
	
	\subsection{Impact of the numerical aperture (NA) and focusing depth on the nonlinear pulse propagation}
	\label{subsec:ImpactNAToPulsePropagation}
	Changing the physical position of the lens itself (i.e. shifting the depth of the focal spot inside the material) and the parameters of the lens itself (i.e. changing the numerical aperture) influences how the beam propagates inside the material, and how the different nonlinear material parameters can act on the propagating beam. For a varying depth and a fixed numerical aperture the difference is visible for a numerical aperture of \num{0.25}, as shown in \Cref{fig:NA025DepthComp}.
	
	\begin{figure*}[!htb]
		\centering
		\hspace*{\fill}%
		\begin{subfigure}[t]{.245\linewidth}
			\includegraphics[width=\linewidth, height=.8\linewidth]{graphs/tikz-pictures/NA25TF0100um1550nm100nJ100fs100umr149t_199201.tikz}
			\caption{$ \lambda=\SI{1550}{\nano\meter} $, NA = \num{0.25}}
			\label{subfig:l1550N025D100}
		\end{subfigure}\hfill%
		\begin{subfigure}[t]{.245\linewidth}
			\includegraphics[width=\linewidth, height=.8\linewidth]{graphs/tikz-pictures/NA25TF0100um1950nm100nJ100fs100umr149t_199201.tikz}
			\caption{$ \lambda=\SI{1950}{\nano\meter} $, NA = \num{0.31}}
			\label{subfig:l1950N025D100}
		\end{subfigure}\hfill%
		\begin{subfigure}[t]{.245\linewidth}
			\includegraphics[width=\linewidth, height=.8\linewidth]{graphs/tikz-pictures/NA25TF0100um2150nm100nJ100fs100umr149t_199201.tikz}
			\caption{$ \lambda=\SI{2150}{\nano\meter} $, NA = \num{0.35}}
			\label{subfig:l2150N025D100}
		\end{subfigure}\hfill%
		\begin{subfigure}[t]{.245\linewidth}
			\includegraphics[width=\linewidth, height=.8\linewidth]{graphs/tikz-pictures/NA25TF0100um2550nm100nJ100fs100umr149t_199201.tikz}
			\caption{$ \lambda=\SI{2550}{\nano\meter} $, NA = \num{0.41}}
			\label{subfig:l2550N025D100}
		\end{subfigure}\hspace*{\fill}%
		
		\hspace*{\fill}%
		\begin{subfigure}[t]{.245\linewidth}
			\includegraphics[width=\linewidth, height=.8\linewidth]{graphs/tikz-pictures/NA25TF0250um1550nm100nJ100fs250umr149t_199201.tikz}
			\caption{$ \lambda=\SI{1550}{\nano\meter} $, NA = \num{0.25}}
			\label{subfig:l1550N025D250}
		\end{subfigure}\hfill%
		\begin{subfigure}[t]{.245\linewidth}
			\includegraphics[width=\linewidth, height=.8\linewidth]{graphs/tikz-pictures/NA25TF0250um1950nm100nJ100fs250umr149t_199201.tikz}
			\caption{$ \lambda=\SI{1950}{\nano\meter} $, NA = \num{0.31}}
			\label{subfig:l1950N025D250}
		\end{subfigure}\hfill%
		\begin{subfigure}[t]{.245\linewidth}
			\includegraphics[width=\linewidth, height=.8\linewidth]{graphs/tikz-pictures/NA25TF0250um2150nm100nJ100fs250umr149t_199201.tikz}
			\caption{$ \lambda=\SI{2150}{\nano\meter} $, NA = \num{0.35}}
			\label{subfig:l2150N025D250}
		\end{subfigure}\hfill%
		\begin{subfigure}[t]{.245\linewidth}
			\includegraphics[width=\linewidth, height=.8\linewidth]{graphs/tikz-pictures/NA25TF0250um2550nm100nJ100fs250umr149t_199201.tikz}
			\caption{$ \lambda=\SI{2550}{\nano\meter} $, NA = \num{0.41}}
			\label{subfig:l2550N025D250}
		\end{subfigure}\hspace*{\fill}%
		
		\hspace*{\fill}%
		\begin{subfigure}[t]{.245\linewidth}
			\includegraphics[width=\linewidth, height=.8\linewidth]{graphs/tikz-pictures/NA85TF0250um1550nm100nJ100fs250umr149t_199201.tikz}
			\caption{$ \lambda=\SI{1550}{\nano\meter} $, NA = \num{0.85}}
			\label{subfig:l1550N085D250}
		\end{subfigure}\hfill%
		\begin{subfigure}[t]{.245\linewidth}
			\includegraphics[width=\linewidth, height=.8\linewidth]{graphs/tikz-pictures/NA85TF0250um1950nm100nJ100fs250umr149t_199201.tikz}
			\caption{$ \lambda=\SI{1950}{\nano\meter} $, NA = \num{1.08}}
			\label{subfig:l1950N085D250}
		\end{subfigure}\hfill%
		\begin{subfigure}[t]{.245\linewidth}
			\includegraphics[width=\linewidth, height=.8\linewidth]{graphs/tikz-pictures/NA85TF0250um2150nm100nJ100fs250umr149t_199201.tikz}
			\caption{$ \lambda=\SI{2150}{\nano\meter} $, NA = \num{1.15}}
			\label{subfig:l2150N085D250}
		\end{subfigure}\hfill%
		\begin{subfigure}[t]{.245\linewidth}
			\includegraphics[width=\linewidth, height=.8\linewidth]{graphs/tikz-pictures/NA85TF0250um2550nm100nJ100fs250umr149t_199201.tikz}
			\caption{$ \lambda=\SI{2550}{\nano\meter} $, NA = \num{1.40}}
			\label{subfig:l2550N085D250}
		\end{subfigure}\hspace*{\fill}%
		\caption{Electric field (Intensity) for $ t_0=\SI{100}{\femto\second} $, $ f_\text{inside}=\SI{100}{\micro\meter} $, $ E_0=\SI{100}{\nano\joule} $ directly at the focal spot ($ z=\SI{100}{\micro\meter} $ for \Cref{subfig:l1550N025D100,subfig:l1950N025D100,subfig:l2150N025D100,subfig:l2550N025D100} and $ z=\SI{250}{\micro\meter} $ for \Cref{subfig:l1550N025D250,subfig:l1950N025D250,subfig:l2150N025D250,subfig:l2550N025D250,subfig:l1550N085D250,subfig:l1950N085D250,subfig:l2150N085D250,subfig:l2550N085D250}). Small variations in the field distribution are visible for \SI{1950}{\nano\meter} (c.f. \Cref{subfig:l1950N025D100,subfig:l1950N025D250}) and \SI{2150}{\nano\meter} (c.f. \Cref{subfig:l2150N025D100,subfig:l2150N025D250}). The intensity field distribution for \SI{1550}{\nano\meter} and \SI{2550}{\nano\meter} does not change, in comparison. In \Cref{subfig:l1550N085D250,subfig:l1950N085D250,subfig:l2150N085D250,subfig:l2550N085D250} the intensity field for an NA of \num{0.85} is displayed in comparison, with similar parameters as \Cref{subfig:l1550N025D250,subfig:l1950N025D250,subfig:l2150N025D250,subfig:l2550N025D250}. Unlike as for a numerical aperture of \num{0.25} the field distribution for a numerical aperture of \num{0.85} does not change significantly when decreasing the focal depth from \SI{250}{\micro\meter} to \SI{100}{\micro\meter}, and therefore was omitted.}
		\label{fig:NA025DepthComp}
	\end{figure*}
	
	Moving the focal spot deeper into the material will change the intensity distribution along the propagation path, such that nonlinear effects, especially self-focusing, can act on the beam for a longer period, and therefore change the resulting intensity field distribution at the focal spot. This behavior is visible especially for $ \lambda=\SI{1950}{\nano\meter},\,\SI{2150}{\nano\meter} $, after those wavelengths experience a stronger self-focusing compared to shorter and longer wavelengths (as shown in \Cref{fig:nPA-Absorption}). This effect is visible especially along the beam propagation axis, as shown in \Cref{fig:SlicesNA085Depth100Energy100Duration100Comp}.
	
	When increasing the numerical aperture from \num{0.25} to \num{0.85} the initial intensity along the propagation path is lower, and therefore nonlinear effects will start closer to the focal spot. Therefore, almost no difference is visible in the structure of the intensity field distribution for an NA of \num{0.85} is visible (c.f. \Cref{subfig:SliceIntensity085100100Comp,subfig:SliceIntensity085100250Comp}), while the pulse focused with a lower NA value changes its structure when changing the focal depth. This demonstrates that larger focusing angles are necessary for processing closer to the surface, where the interaction length with nonlinear effects will be reduced around the focal spot compared to lower focusing angles.
	
	\subsection{Spatio-temporal pulse behavior depending on pulse wavelength and -energy}
	\label{subsec:SpatioTemporalPulseWLDependence}

	\begin{figure}[htb]
		\centering
		\hspace*{\fill}%
		\begin{subfigure}[t]{\linewidth}
			\centering
			\includegraphics[width=.9\linewidth, height=.4\linewidth]{images/Images_Vladimir/fig3250_NA0_25_cut.tikz}
			\caption{$ z=\SI{300}{\micro\meter} $, $ \text{NA}=\num{0.25} $}
			\label{subfig:longwavelength}
		\end{subfigure}\hfill
		\begin{subfigure}[t]{\linewidth}
			\centering
			\includegraphics[width=.9\linewidth, height=.4\linewidth]{images/Images_Vladimir/fig13_cut.tikz}
			\caption{$ z=\SI{270}{\micro\meter} $, $ \text{NA}=\num{0.85} $}
			\label{subfig:longwavelengthTBD}
		\end{subfigure}\hspace*{\fill}%
		\caption{Contour plot of the intensity at $ z=\SI{270}{\micro\meter} $ in \Cref{subfig:longwavelengthTBD} and $ z=\SI{300}{\micro\meter} $ in \Cref{subfig:longwavelength} inside the medium. The numerical aperture is \num{0.25} in \Cref{subfig:longwavelength}, and \num{0.85} in \Cref{subfig:longwavelengthTBD}. The input pulse parameters (following \Cref{equ:InitialConditionsHelmholtzDescription}) are $ P_0=\SI{0.88}{\mega\watt} $, $ f=\SI{500}{\micro\meter} $, NA = \num{0.25}, $ w_0=\SI{4}{\micro\meter}/w_0=\SI{1.2}{\micro\meter} $ for \Cref{subfig:longwavelength}/\Cref{subfig:longwavelengthTBD}, respectively, $ \lambda=\SI{3250}{\nano\meter} $, and $ t_0=\SI{100}{\femto\second} $. 
		\label{fig:longwavelengthFigure}
		}
	\end{figure}
	
	The spatio-temporal splitting described in \Cref{subsec:SpatiotemporalStructure} and shown in \Cref{subfig:z1473,subfig:z1578,subfig:z1894,subfig:z2000} does depend on several different factors such as pulse wavelength, pulse energy, pulse duration and the focusing optics. While the first parameter influences the value of the material parameters (as seen in \Cref{tab:PhysParamsExample}), the latter three influence how the intensity of the pulse will behave while travelling through the material. A higher intensity will result in both earlier self-focusing and a faster onset of multi-photon absorption, which in turn are both influenced by the value of the material parameters set by the wavelength. As a first comparison the intensity of a pulse with $ t_0=\SI{100}{\femto\second} $ and $ E_0=\SI{100}{\nano\joule} $ is shown in \Cref{subfig:l1550N085D250,subfig:l1950N085D250,subfig:l2150N085D250,subfig:l2550N085D250}, when focused with a target NA of \num{0.85} (ref. \Cref{tab:AdjustedNAValues}) and the focus point set to \SI{250}{\micro\meter} below the surface of \ce{Si}. In addition to the spatio-temporal representation of the field shown in \Cref{subfig:l1550N085D250,subfig:l1950N085D250,subfig:l2150N085D250,subfig:l2550N085D250} corresponding slices taken at $ r=\SI{0}{\micro\meter} $ for both the temporal intensity and the generated free carrier density are shown in \Cref{subfig:SliceFCDensity085100100Comp,subfig:SliceIntensity085100100Comp} for a focus depth of \SI{100}{\micro\meter} and in \Cref{subfig:SliceFCDensity085100250Comp,subfig:SliceIntensity085100250Comp} for a focus depth of \SI{250}{\micro\meter}.
	
	The effect of multi-photon absorption and plasma diffraction can clearly be seen for all pulse wavelengths. The trailing part of the pulse has been absorbed almost completely, resulting in a shift of the peak intensity towards the leading edge. In addition the self-focusing effect introduces oscillations, especially for \SI{1950}{\nano\meter} and \SI{2150}{\nano\meter}.
	
	Here, it is also worth to mention the intensity distribution over the wavelengths at the focal spot. For the linear propagation without absorption or diffraction the intensities should be equal at the focal spot, but even though the self-focusing effect is stronger at \SI{1950}{\nano\meter} and \SI{2150}{\nano\meter} compared to the shorter and longer wavelengths, the peak intensity follows the absorption curve shown in \Cref{fig:CurrentComparisonMPA} at $ I_0\geq\SI{1e15}{\watt\per\square\meter} $, with the shortest wavelength having the highest peak intensity. Nevertheless, the difference between the peak intensities is negligible, with the peak intensity at \SI{2550}{\nano\meter} only $ \approx\num{1.7} $ times smaller than the peak intensity at \SI{1550}{\nano\meter}. In comparison to this difference in intensity the density of the generated free carriers is roughly comparable for all wavelengths independently of the depth of the focus point (as shown in \Cref{subfig:SliceFCDensity085100100Comp,subfig:SliceFCDensity085100250Comp}), with the generated density being slightly higher for \SI{2550}{\nano\meter} by $ \approx\SI{8}{\percent} $. This indicates that the absorbed energy going into the generation of free carriers is roughly the same, independently of the wavelength.
	
	In comparison to a lens with an NA of \num{0.85} one also notices that the field is longer in time (especially visible in the slices shown in \Cref{subfig:SliceIntensity085100100Comp,subfig:SliceIntensity085100250Comp,subfig:SliceIntensity025100100Comp,subfig:SliceIntensity025100250Comp}). For a numerical aperture of \num{0.25} the pulse is stretched to over $ \approx\SI{800}{\femto\second} $, while for a numerical aperture of \num{0.85} the pulse is only stretched to \SIrange[range-units=single, range-phrase=\text{ to }]{100}{200}{\femto\second}, depending on the observed wavelength, and thereby much closer to the original pulse duration.
	
	For long wavelengths, the spherical aberrations decay so that a field concentrates in a comparatively symmetrical spot at some propagation distance inside a medium (see \Cref{fig:longwavelengthFigure}). As one can see from~\Cref{fig:nPA-Absorption,fig:3dNFOM}, the two-photon absorption process decreases towards this wavelength range so that a relative contribution of the Kerr-effect increases. Thus, a plasma influence on a beam focusing destabilization and a damage threshold decays. As a result, a non-destructive modification of material becomes possible. One must note that the weakening of plasma contribution breaks the spatial-time relation mentioned above. Namely, a beam squeezes mainly in spatial dimension than in the time one for the large NA (compare~\Cref{fig:longwavelengthFigure}). Our analysis demonstrates that such an effect could cause a spatial-time splitting at some propagation distances, which requires control of initial beam parameters.
	
	\subsection{Correlation between the pulse duration, the pulse wavelength, the pulse energy and the peak intensity/peak free carrier density/absorbed energy}
	\label{sec:CorrelationPulseDurationPulseIntensity}
	
\begin{figure*}[htb]
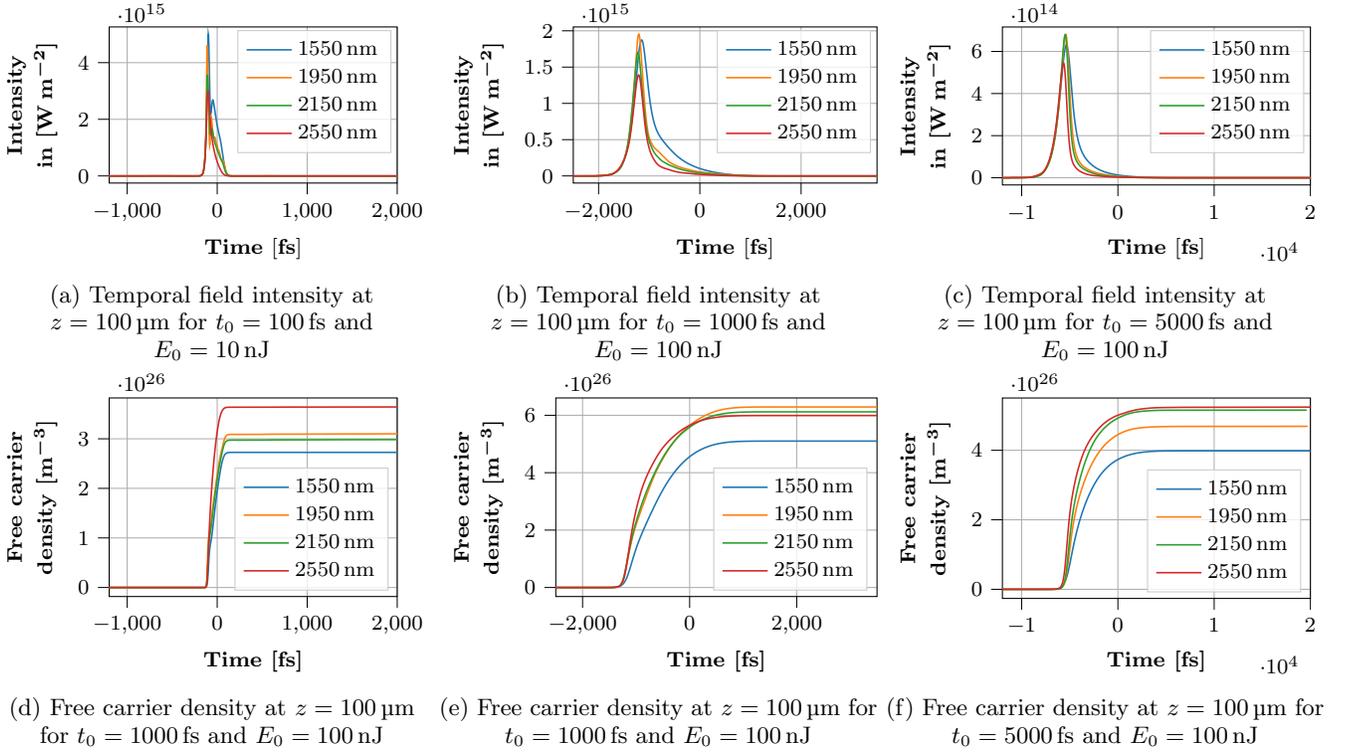

		\centering
		\hspace*{\fill}%
		\begin{subfigure}[t]{.325\linewidth}
			\centering
			\includegraphics[width=\linewidth, height=.6\linewidth]{graphs/tikz-slices/NA85TS0um10nJ100fs100um.tikz}
			\caption{Temporal field intensity at $ z=\SI{100}{\micro\meter} $ for $ t_0=\SI{100}{\femto\second} $ and $ E_0=\SI{10}{\nano\joule} $}
			\label{subfig:SliceIntensity08510100}
		\end{subfigure}\hfill%
		\begin{subfigure}[t]{.325\linewidth}
			\centering
			\includegraphics[width=\linewidth, height=.6\linewidth]{graphs/tikz-slices/NA85TS0um100nJ1000fs100um.tikz}
			\caption{Temporal field intensity at $ z=\SI{100}{\micro\meter} $ for $ t_0=\SI{1000}{\femto\second} $ and $ E_0=\SI{100}{\nano\joule} $}
			\label{subfig:SliceIntensity0851001000}
		\end{subfigure}\hfill%
		\begin{subfigure}[t]{.325\linewidth}
			\centering
			\includegraphics[width=\linewidth, height=.6\linewidth]{graphs/tikz-slices/NA85TS0um100nJ5000fs100um.tikz}
			\caption{Temporal field intensity at $ z=\SI{100}{\micro\meter} $ for $ t_0=\SI{5000}{\femto\second} $ and $ E_0=\SI{100}{\nano\joule} $}
			\label{subfig:SliceIntensity0851005000}
		\end{subfigure}\hspace*{\fill}%
		
		\hspace*{\fill}%
		\begin{subfigure}[t]{.325\linewidth}
			\centering
			\includegraphics[width=\linewidth, height=.7\linewidth]{graphs/tikz-slices/NA85DS0um10nJ100fs100um.tikz}
			\caption{Free carrier density at $ z=\SI{100}{\micro\meter} $ for $ t_0=\SI{1000}{\femto\second} $ and $ E_0=\SI{100}{\nano\joule} $}
			\label{subfig:SliceFCDensity08510100}
		\end{subfigure}\hfill%
		\begin{subfigure}[t]{.325\linewidth}
			\centering
			\includegraphics[width=\linewidth, height=.7\linewidth]{graphs/tikz-slices/NA85DS0um100nJ1000fs100um.tikz}
			\caption{Free carrier density at $ z=\SI{100}{\micro\meter} $ for $ t_0=\SI{1000}{\femto\second} $ and $ E_0=\SI{100}{\nano\joule} $}
			\label{subfig:SliceFCDensity0851001000}
		\end{subfigure}\hfill%
		\begin{subfigure}[t]{.325\linewidth}
			\centering
			\includegraphics[width=\linewidth, height=.7\linewidth]{graphs/tikz-slices/NA85DS0um100nJ5000fs100um.tikz}
			\caption{Free carrier density at $ z=\SI{100}{\micro\meter} $ for $ t_0=\SI{5000}{\femto\second} $ and $ E_0=\SI{100}{\nano\joule} $}
			\label{subfig:SliceFCDensity0851005000}
		\end{subfigure}\hspace*{\fill}%
		\caption{Slice of the free-carrier density and the temporal field intensity for a pulse with $ E_0=\SI{10}{ \nano\joule}/\SI{100}{\nano\joule} $ and $ t_0=\SI{100}{\femto\second}/\SI{1000}{\femto\second}/\SI{5000}{\femto\second} $, focused using a lens with an NA of \num{0.85}/\num{1.07}/\num{1.18}/\num{1.40} and a focus depth of \SI{100}{\micro\meter}}
		\label{fig:SlicesNA085Depth100Energy10100Duration10010005000}
	\end{figure*}

	As shown in \Cref{fig:CurrentComparisonMPA} the absorption of the pulse largely depends on the pulse intensity. To reduce the intensity of the pulse there are several options: Increasing the focal spot size, decreasing the pulse energy or increasing the pulse duration. With the size of the modifications in \ce{Si} as the main goal the first option is not ideal, and therefore either the pulse energy has to be decreased, or the pulse duration increased. The effectiveness of both approaches is compared in \Cref{fig:SlicesNA085Depth100Energy10100Duration10010005000}.
	
	The corresponding fields (for \SI{1550}{\nano\meter}) are shown in \Cref{fig:SlicesNA085Depth100Energy10100Duration10010005000}. Several effects can be seen here, and are worth mentioning:
	\begin{itemize}
		\item A longer pulse looses a larger amount of energy, leading to the larger shift of the intensity peak towards the leading edge (c.f. \Cref{subfig:SliceIntensity08510100,subfig:SliceIntensity0851001000})
		\item While the pulse with $ \lambda=\SI{1550}{\nano\meter} $ generates the highest intensity at the focal spot for the pulses with $ t_0=\SI{100}{\femto\second} $ (c.f. \Cref{subfig:SliceIntensity085100100Comp,subfig:SliceIntensity085100250Comp,subfig:SliceIntensity08510100}). For longer pulse durations ($ \geq\SI{1000}{\femto\second} $) higher intensities will be generated at longer wavelengths.
		\item Even though the pulse energy is one order of magnitude lower in \Cref{subfig:SliceIntensity08510100} than in \Cref{subfig:SliceIntensity085100100Comp} the peak intensity value is approximately equal, indicating intensity clamping due to high absorption and diffraction.
		\item The generated plasma density for a pulse with $ t_0=\SI{1000}{\femto\second} $ is \SI{50}{\percent} larger compared to a pulse with $ t_0=\SI{100}{\femto\second} $, as shown in \Cref{subfig:SliceFCDensity085100100Comp,subfig:SliceFCDensity0851001000}, if both pulses have an energy of $ E_0=\SI{100}{\nano\joule} $. This indicates a better energy deposition of the pulse energy into the material for longer pulses
		\item While the pulse with $ \lambda=\SI{2550}{\nano\meter} $ at $ t_0=\SI{100}{\femto\second} $ could generate the highest free-carrier density, this role is taken over by the pulse with $ \lambda=\SI{1950}{\nano\meter} $ for $ t_0=\SI{1000}{\femto\second} $, with the pulses at $ \lambda=\SI{2150}{\nano\meter} $ and $ \lambda=\SI{2550}{\nano\meter} $ on second position.
	\end{itemize}
	
	\begin{figure}[htb]
		\centering
		\begin{subfigure}[t]{.9\linewidth}
			\centering
			\includegraphics[width=\linewidth, height=.5\linewidth]{graphs/tikz-pictures/NA85TF0100um1550nm10nJ100fs100umr149t_199201.tikz}
			\caption{$ t_0=\SI{100}{\femto\second} $, $ E_0=\SI{10}{\nano\joule} $}
			\label{subfig:NA085Depth100Energy10Duration100}
		\end{subfigure}
		\hfill%
		\begin{subfigure}[t]{.9\linewidth}
			\centering
			\includegraphics[width=\linewidth, height=.5\linewidth]{graphs/tikz-pictures/NA85TF0100um1550nm100nJ1000fs100umr149t_20011500.tikz}
			\caption{$ t_0=\SI{1000}{\femto\second} $, $ E_0=\SI{100}{\nano\joule} $}
			\label{subfig:NA085Depth100Energy100Duration1000}
		\end{subfigure}%
		\caption{Electric field intensity for pulses with $ \lambda=\SI{1550}{\nano\meter} $, focused \SI{100}{\micro\meter} below the surface of \ce{Si}, with ($ t_0=\SI{100}{\femto\second} $/$ E_0=\SI{10}{\nano\joule} $) in \Cref{subfig:NA085Depth100Energy10Duration100} and ($ t_0=\SI{1000}{\femto\second} $/$ E_0=\SI{100}{\nano\joule} $) in \Cref{subfig:NA085Depth100Energy100Duration1000}. The numerical aperture for both pulses is \num{0.85}}
		\label{fig:NA085Depth100Energy10100Duration1001000}
	\end{figure}
	
	This indicates that pulses longer than \SI{100}{\femto\second} are necessary to efficiently transfer energy from the pulse into the material. For a pulse duration of \SI{1000}{\femto\second} exemplary results are displayed in \Cref{subfig:SliceIntensity0851001000,subfig:SliceFCDensity0851001000}, and for a pulse duration of \SI{5000}{\femto\second} those results are displayed in \Cref{subfig:SliceIntensity0851005000,subfig:SliceFCDensity0851005000}. Several major differences can be noted for a pulse duration of \SI{5000}{\femto\second} in comparison with the data in \Cref{subfig:SliceIntensity0851001000} (for $ t_0=\SI{1000}{\femto\second} $): 
	\begin{itemize}
		\item The pulses at $ \lambda=\SI{1950}{\nano\meter} $ and $ \lambda=\SI{2150}{\nano\meter} $ now share the peak intensity value, compared with the pulses with a shorter and longer wavelength 
		\item The peak intensity for $ t_0=\SI{5000}{\femto\second} $ is decreased by $ \approx\SI{65}{\percent} $ compared to $ t_0=\SI{1000}{\femto\second} $
		\item The peak density decreased by $ \approx\SI{20}{\percent} $ compared to $ t_0=\SI{1000}{\femto\second} $, but is still higher by $ \approx\SI{25}{\percent} $ compared to a pulse duration of \SI{100}{\femto\second}
	\end{itemize}
	
	As it can be seen in \Cref{fig:SlicesNA085Depth100Energy100Duration100Comp,subfig:SliceIntensity0851005000,subfig:SliceFCDensity0851005000,subfig:SliceFCDensity08510100,subfig:SliceIntensity08510100,subfig:SliceIntensity0851001000,subfig:SliceFCDensity0851001000} the peak intensity and the carrier density vary depending on pulse duration, pulse energy and pulse wavelength if the focal depth and focusing angle is kept at a constant value. Following those graphs for a focusing depth of \SI{100}{\micro\meter} and a numerical aperture of \num{0.85}/\num{1.07}/\num{1.18}/\num{1.40} for the different wavelengths, correspondingly, it can be seen that the maximum intensity can be achieved for short pulses with high energies. Meanwhile, the free carrier density does not follow the same pattern. Instead, very short and very long pulses generate a lower amount of free carriers compared to pulses with a duration of $ \approx\SIrange[range-units=single, range-phrase=\text{ to }]{600}{800}{\femto\second} $. The initial energy of the pulse, on the other hand, directly correlates with the generated amount of free carriers. A similar pattern can be seen for the other parameters we investigated. Regardless of focus depth or numerical aperture the maximum intensity could usually be reached for the shortest possible pulse duration, combined with the highest possible pulse energy. The maximum free carrier density, though, correlates is maximized for a medium pulse duration.
	
	The results shown in the figures above (\Cref{subfig:SliceFCDensity08510100,subfig:SliceFCDensity08510100,subfig:SliceFCDensity0851005000}) indicate that there is an optimum of the pulse parameters for the optimum energy deposition into the material. Even though a shorter pulse with $ t_0=\SI{100}{\femto\second} $ can reach a higher intensity compared to longer pulses at the same pulse energy it is not able to transfer the energy into the material as efficiently as pulses with $ t_0=\SI{1000}{\femto\second} $ or \SI{5000}{\femto\second}, assumed the efficiency of the energy transfer is determined by the amount of excited free carriers.
	
	\begin{figure*}[!htb]
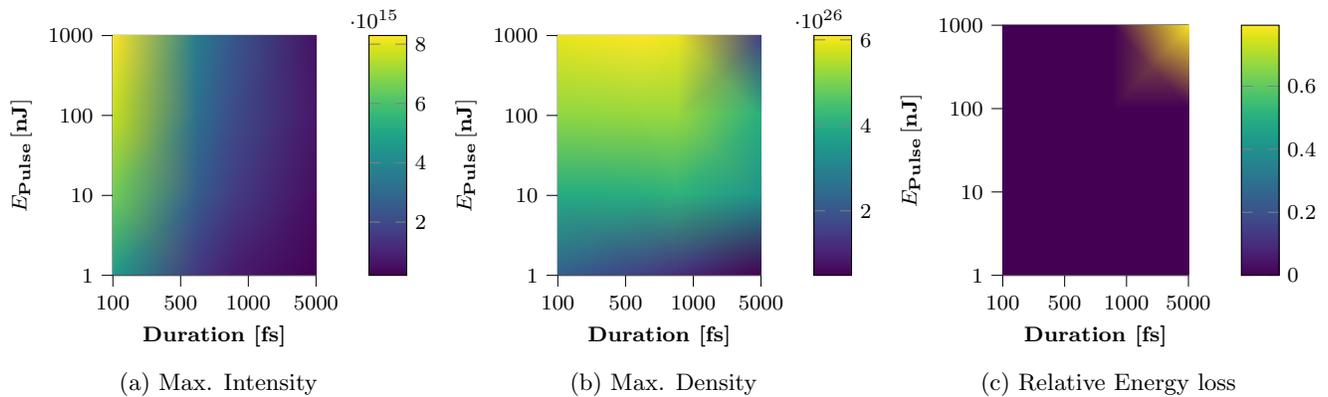

		\centering
		\hspace*{\fill}%
		\begin{subfigure}[t]{.32\linewidth}
			\centering
			\includegraphics[width=\linewidth, height=.8\linewidth]{graphs/pulse-surfaces/NA_085_Intensity_1550_nm_100_um.tikz}
			\caption{Max. Intensity}
			\label{subfig:MaxIntensityPeak}
		\end{subfigure}\hfill%
		\begin{subfigure}[t]{.32\linewidth}
			\centering
			\includegraphics[width=\linewidth, height=.8\linewidth]{graphs/pulse-surfaces/NA_085_Density_1550_nm_100_um.tikz}
			\caption{Max. Density}
			\label{subfig:MaxDensityPeak}
		\end{subfigure}\hfill%
		\begin{subfigure}[t]{.32\linewidth}
			\centering
			\includegraphics[width=\linewidth, height=.8\linewidth]{graphs/pulse-surfaces/NA_085_RelEnergyDrop_1550_nm_100_um.tikz}
			\caption{Relative Energy loss}
			\label{subfig:RelEnergyLoss}
		\end{subfigure}\hspace*{\fill}%
		\caption{Influence of the pulse parameters energy and duration at a constant wavelength of \SI{1550}{\nano\meter}, numerical aperture of \num{0.85} and focus depth of \SI{100}{\micro\meter} below the surface. A shorter pulse duration is directly correlated with a higher peak intensity (c.f. \Cref{subfig:MaxIntensityPeak}), while a pulse with a duration between \SIrange[range-units=single, range-phrase=\text{ to }]{500}{1000}{\femto\second} results in an increased free carrier density compared to longer and shorter pulse durations (c.f. \Cref{subfig:MaxDensityPeak}). In \Cref{subfig:RelEnergyLoss} the correlation between pulse duration, pulse energy and the relative energy loss within $ \SI{\pm5}{\percent} $ around the focal spot (i.e. within \SIrange[range-units=single, range-phrase=\text{ to }]{95}{105}{\micro\meter} in the material) are shown. Here the longest pulse durations together with the highest pulse energy give the maximum relative energy loss at the focal spot. The colorbar in \Cref{subfig:MaxIntensityPeak} gives the intensity in $ \left[\si{\watt\per\square\meter}\right] $, in \Cref{subfig:MaxDensityPeak} the density of the free carriers in $ \left[\si{\per\cubic\meter}\right] $ and in \Cref{subfig:RelEnergyLoss} the relative energy loss in $ \left[\si{\percent}\right] $}
		\label{fig:EfficiencyComparisonFigures}
	\end{figure*}
	
	\subsection{Correlation between pulse parameters and the permanent material modification}
	\label{subsec:CorrPulseParametersMaterialModification}
	To process material it is always necessary to introduce permanent modifications in the material. As stated above in \Cref{subsec:MechanismOfSubSurfaceModificationSiTheory} there are several different ways to define the lowest threshold for material modification. Following \Citet{Werner2019,Chambonneau2021} we do not use the critical plasma density as threshold, but rather the deposited energy. This energy can be derived from the excited electrons by using \Cref{equ:AbsorbedEnergy} and the assumption of full thermalization of all excited electrons. This equation together with the specific heat capacity of \ce{Si} (\SI{0.71}{\joule\per\gram\per\kelvin}=\SI{1.65}{\joule\per\cubic\centi\meter\per\kelvin}~\cite{Chase1998}) and the latent heat of fusion of \ce{Si} ($\SI{50.55}{\kilo\joule\per\mol}=\SI{1.80}{\kilo\joule\per\gram}$~\cite{Chase1998}) gives the temperature change (and the possibility for a phase change). Using the results for the achieved free carrier density, displayed in \Cref{subfig:SliceFCDensity025100100Comp,subfig:SliceFCDensity025100250Comp,subfig:SliceFCDensity085100100Comp,subfig:SliceFCDensity085100250Comp,subfig:MaxDensityPeak} and ranging between \SIrange[range-units=single, range-phrase=\text{ to }]{2e26}{6e26}{\per\cubic\meter}, one obtains absorbed energies between $\approx\SIrange[range-units=single, range-phrase=\text{ to }]{72}{215}{\joule\per\cubic\centi\meter}$. Absorbing these energies will lead to a temperature increase of $ \approx\SIrange[range-units=single, range-phrase=\text{ to }]{44}{130}{\kelvin} $. This is not sufficient for increasing the temperature of \ce{Si} from room temperature to the melting temperature, but sufficient to expand the heated spot slightly (with $\Delta V/V=\SI{9e-6}{\per\kelvin}$~\cite{Chase1998}). This local expansion can lead to a permanent local change of the refractive index (even without melting), which then in turn can be detected with optical methods. 
	
	This direct correlation between the generated free carrier density and the resulting change in temperature also confirms that the modification threshold does not only depend on the wavelength, but rather on the interplay between several different pulse parameters including the pulse wavelength. Especially for longer pulse durations and large focusing angles several different wavelengths can achieve approximately the same energy deposit (c.f. \Cref{subfig:SliceFCDensity085100100Comp,subfig:SliceFCDensity085100250Comp,subfig:SliceFCDensity08510100,subfig:SliceFCDensity025100100Comp,subfig:SliceFCDensity025100250Comp,subfig:SliceFCDensity0851001000,subfig:SliceFCDensity0851005000}). In that case pulse parameters other than the wavelength have a larger influence, such as initial pulse energy, pulse duration and focusing angle.

	\section{Conclusion and outlook}
	\label{sec:Conclusion}
	We investigated the behavior of a single pulse propagating through a material on example of \ce{Si}, with specific focus on achieving particularly fine processing at low pulse energies. The influence of the wavelength, the nonlinear multi-photon absorption, the third-order nonlinearity, the focusing angle, the focal depth, the pulse energy and the pulse duration were considered, with several important conclusions:
	\begin{itemize}
		\item We confirmed that using wavelengths between \SIrange[range-units=single, range-phrase=\text{ to }]{2000}{2200}{\nano\meter} is beneficial for maximizing the absorption of light and generation of free carriers in \ce{Si} even at high focusing angles compared to shorter and longer wavelengths at the optimum pulse duration. This follows from the predicted curve in the nonlinear figure of merit, which has two distinct maxima at $\approx\SI{2100}{\nano\meter}$ and $\approx\SI{3200}{\nano\meter}$ (as shown in \Cref{fig:3dNFOM}).
		\item The importance of nonlinear effects inside \ce{Si} for the propagation is larger for smaller NA values. The larger the focusing angle, the less influence self-focusing and multi-photon absorption has on the propagating pulse before the focal spot. This increases the repeatability of modification generation.
		\item The effect of nonlinear spherical aberrations or \actualquote{light ring generation} was investigated. As it was shown, this effect degrades close to the maximum of the NFOM. The longer the wavelength, the better the quality of the processed structures.
		\item The wavelength-dependent effects related to the Kerr effect and the multi-photon absorption defined by NFOM illustrate the different mechanisms of material processing. Namely, the plasma generation due to two- and three-photon absorption enhanced at $\approx\SI{2100}{\nano\meter}$ by the Kerr-effect would decrease the modification threshold.
		\item We could show that the field distribution at the focal spot only varies slightly when changing the focal depth. Increasing the focusing angle decreases changes, i.e. for larger numerical apertures the distribution of the electric field changes less when changing the target depth compared to smaller numerical apertures
		\item It could be shown that there is an optimum pulse duration, located between \SIrange[range-units=single, range-phrase=\text{ to }]{600}{900}{\femto\second}, explained by the necessity to optimize the spectral overlap between the pulse and the nonlinear figure of merit.
		\item Based on the two independent numerical approaches, we show that a nonparaxial beam structure under a plasma contribution could cause a spatio-temporal asymmetry and pulse splitting even for a comparatively small NA. Advantages of these numerical approaches are compared.
	\end{itemize}
	This work has thus demonstrated the importance of the careful control of laser parameters and in particular wavelength, pulse duration and energy within a large window for optimization of the energy transfer from the light pulse to the material at the focal spot. It sheds light on the different physical phenomena acting on the pulse, including the Kerr-effect, and how they can be used for optimizing the energy transfer. Further optimization and analysis can involve several directions. For example, wavelengths larger than \SI{3000}{\nano\meter} can be investigated. As shown in \Cref{fig:3dNFOM,fig:NonFig1e13InText} there is a transition at $ \approx\SI{3300}{\nano\meter} $ from the three-photon absorption to the four-photon absorption. Placing a pulse within those two absorption regions (for example at \SI{3250}{\nano\meter}) would have the advantage of a decreased absorption up close to the focal spot (c.f. \Cref{fig:CurrentComparisonMPA}), but would loose the additional benefit of increased self-focusing, as the pulses at \SIrange[range-units=single, range-phrase=\text{ to }]{1950}{2150}{\nano\meter} can use.
	
	Finally, the wavelength dependence and the impact of tunnel ionization versus multi-photon ionization and avalanche ionization can be studied and compared in a future work.
	
	\section*{Acknowledgements}
	The authors would like to thank Cord Arnold (Lund University, Lund) and Andreas Erbe (NTNU, Trondheim) for the useful discussions. We also would like to thank Miroslav Kolesik (The University of Arizona, Arizona), Jeffrey Brown (Wisconsin Lutheran College, Milwaukee), Arnaud Couairon (CPHT, Ecole Polytechnique, Palaiseau) and Vladimir Fedorov (Texas A\&M University at Qatar) for the very fruitful discussions about the UPPE-model. RR and ITS acknowledge financial support of the Norwegian Research Council (NFR) projects \#255003 (KerfLessSi), \#326503 (MIR), \#303347 (UNLOCK) and ATLA Lasers AS. VK acknowledges the support from the European Union Horizon 2020 research and innovation program under the Marie Sk{\l}odowska-Curie grant No. 713694 (MULTIPLY) and the ERC Advanced Grant No. 740355 (STEMS).
	
	\section*{Disclosures}
	The authors declare no conflicts of interest.
	\appendix
	\section{Initial conditions for the electric field}
	\label{subsec:InitialConditionsField}
	To model the pulse as close as possible to existing experimental conditions, a Gaussian shape in time and space was chosen for the incoming pulse.
	\begin{equation}\label{equ:PulseEquation}
		\begin{split}
			E(r, t) &=\\
			\sqrt{\frac{P_0}{\pi w^2}}&\cdot\exp\left(-\left(\frac{r}{w}\right)^2-\frac{ik_0r^2}{2f}-i\phi - \left(\frac{\Delta t}{T}\right)^2-i\omega_0t\right).
		\end{split}
	\end{equation}
	Here, $r$ gives the position in space perpendicular to the propagation axis, $w$ the beam radius, $ k_0 $ the wave number in vacuum, $ f $ the focal length and $ c $ the speed of light in vacuum. $ z_d=t_0^2/(2\cdot\text{GVD}) $ is the dispersion length, with \textit{GVD} the group velocity dispersion (typically expressed in \si{\femto\second\squared\per\meter}, and labelled as $ \beta_2 $). This value is defined as $\beta_2 = 1/c\left(2d/d\omega\, n(\omega)+\omega d^2/d\omega^2n(\omega)\right)$ based on the pulse frequency $ \omega $ and the refractive index $ n(\omega) $. The pulse broadening due to dispersion is given by $ T=t_0\sqrt{1+\left(z/z_d\right)^2} $, and the phase shift $ \phi $ is defined as $ \phi=\arctan\left(-\frac{c\cdot t}{z_d}\right) $. Finally, the phase shift introduced by the focusing lens ($\Delta t$ in \Cref{equ:PulseEquation}) is defined as $ \Delta t=t + \frac{r^2}{2c(f-z)} $.
	
	In \Cref{equ:Helmholz} we use somewhat more general initial conditions for generality. These conditions correspond to the dynamics of the Gaussian beam in the paraxial approximation:
	\begin{equation}
		\begin{split}
			\frac{\partial E(z = 0, t)}{\partial z} &= 0
		\end{split}
	\end{equation}

	\begin{align}
		\begin{split}
			\Re{\left(E(z=0, t)\right)} &= \sqrt{\frac{P_0}{\pi w^2}}\exp\left(\frac{-r^2}{w^2}\right)\sech\left(\frac{t}{t_0}\right)\\
			&\qquad\cdot\cos\left(\psi-\frac{k_0r^2}{2R}\right)
		\end{split}\\
		\begin{split}
			\Im{\left(E(z=0, t)\right)} &= \sqrt{\frac{P_0}{\pi w^2}}\exp\left(\frac{-r^2}{w^2}\right)\sech\left(\frac{t}{t_0}\right)\\
			&\qquad\cdot\cos\left(\psi-\frac{k_0r^2}{2R}\right)
		\end{split}
	\end{align}
	with the parameters
	\begin{equation}\label{equ:InitialConditionsHelmholtzDescription}
		\begin{split}
			R=f\cdot\left(1+\left(\frac{z_\text{R}}{2}\right)^2\right)\text{, }z_\text{R}&=\frac{w_0}{\pi\text{ NA}}\text{, }w_0 = \frac{\lambda}{\pi\text{ NA}}\\
			w=w_0\sqrt{1+\left(\frac{f}{z_\text{R}}\right)^2}&\text{, }\psi=\arctan\left(\frac{f}{z_\text{R}}\right).
		\end{split}
	\end{equation}
	Of course, real dynamics will differ from the Gaussian one due to nonlinear effects and non-paraxiality. Here, $ P_0 $ is the initial power (as in \Cref{equ:PulseEquation}), $ w $ is the initial beam size, $ w_0 $ is the waist size, $ R $ is the wave front curvature, $ z_\text{R} $ is the Rayleigh length, and $ \psi $ is the Gouy phase. The value of $ \Re{\left(E(z=0,t)\right)^2} + \Im{\left(E(z=0,t)\right)^2} $ has a sense of intensity.
	
	\section{Distribution of the temporal field and density along the propagation axis}
	\label{sec_app:DistributionFieldAlongzAxis}
	The temporal electric field and the distribution of the generated free carriers can not only be displayed as contour plot (as shown in \Cref{fig:NA025DepthComp}), but also observed along the propagation axis. This has the advantage that slight variations in field intensity or carrier density between different focal depths or different numerical apertures are clearer, while only covering a single line along the propagation axis. Still, this also allows a direct overlap of the distribution of field or carrier density, which makes a comparison easier, and which is not possible for contour plots. The slices for the temporal field for a numerical aperture of \num{0.25} and a focal depth of \SI{100}{\micro\meter} (corresponding to \Cref{subfig:l1550N025D100,subfig:l1950N025D100,subfig:l2150N025D100,subfig:l2550N025D100}) are shown in \Cref{subfig:SliceIntensity025100100Comp}, and for a focal depth of \SI{250}{\micro\meter} at the same numerical aperture (corresponding to \Cref{subfig:l1550N025D250,subfig:l1950N025D250,subfig:l2150N025D250,subfig:l2550N025D250}) are shown in \Cref{subfig:SliceIntensity025100250Comp}. For a numerical aperture of \num{0.85} and a focal depth of \SI{100}{\micro\meter} (corresponding to \Cref{subfig:l1550N085D250,subfig:l1950N085D250,subfig:l2150N085D250,subfig:l2550N085D250}) the temporal intensity is displayed in \Cref{subfig:SliceIntensity085100100Comp}, and for a focal depth of \SI{250}{\micro\meter} (without corresponding figures in \Cref{fig:NA025DepthComp}) the intensity is displayed in \Cref{subfig:SliceIntensity085100250Comp}. The density of the excited free carriers corresponding to \Cref{subfig:SliceIntensity025100100Comp,subfig:SliceIntensity025100250Comp,subfig:SliceIntensity085100100Comp,subfig:SliceIntensity085100250Comp} is displayed in \Cref{subfig:SliceFCDensity025100100Comp,subfig:SliceFCDensity025100250Comp,subfig:SliceFCDensity085100100Comp,subfig:SliceFCDensity085100250Comp}.

	\begin{figure}[!htb]
		\centering
		\begin{subfigure}[t]{.9\linewidth}
			\includegraphics[width=\linewidth, height=.4\linewidth]{graphs/tikz-slices/NA25TS0um100nJ100fs100um.tikz}
			\caption{$ z=\SI{100}{\micro\meter} $, $ \text{NA} = \num{0.25} $}
			\label{subfig:SliceIntensity025100100Comp}
		\end{subfigure}\hfill%
	
		\begin{subfigure}[t]{.9\linewidth}
			\includegraphics[width=\linewidth, height=.4\linewidth]{graphs/tikz-slices/NA25TS0um100nJ100fs250um.tikz}
			\caption{$ z=\SI{250}{\micro\meter} $, $ \text{NA} = \num{0.25} $}
			\label{subfig:SliceIntensity025100250Comp}
		\end{subfigure}
	\end{figure}

	\begin{figure*}[!htb]
		\ContinuedFloat
		\hspace*{\fill}%
		\begin{subfigure}[t]{.45\linewidth}
			\includegraphics[width=\linewidth, height=.4\linewidth]{graphs/tikz-slices/NA85TS0um100nJ100fs100um.tikz}
			\caption{$ z=\SI{100}{\micro\meter} $, $ \text{NA} = \num{0.85} $}
			\label{subfig:SliceIntensity085100100Comp}
		\end{subfigure}\hfill%
		\begin{subfigure}[t]{.45\linewidth}
			\includegraphics[width=\linewidth, height=.4\linewidth]{graphs/tikz-slices/NA85TS0um100nJ100fs250um.tikz}
			\caption{$ z=\SI{250}{\micro\meter} $, $ \text{NA} = \num{0.85} $}
			\label{subfig:SliceIntensity085100250Comp}
		\end{subfigure}\hspace*{\fill}%
		\caption{Slice of the temporal field intensity for $ t_0=\SI{100}{\femto\second} $, $ f_\text{inside}=\SI{100}{\micro\meter} $/$f_\text{inside} =\SI{250}{\micro\meter} $, $ E_0=\SI{100}{\nano\joule} $ focused with a lens with an NA = \num{0.25}/\num{0.31}/\num{0.35}/\num{0.41} in \Cref{subfig:SliceIntensity025100100Comp,subfig:SliceIntensity025100250Comp} and an NA = \num{0.85}/\num{1.07}/\num{1.18}/\num{1.40} in \Cref{subfig:SliceIntensity085100100Comp,subfig:SliceIntensity085100250Comp} directly at the focal spot ($ z=\SI{100}{\micro\meter} $/$ z=\SI{250}{\micro\meter} $). The intensity is given in $ \left[\si{\watt\per\square\meter}\right] $}
		\label{fig:SlicesNA085Depth100Energy100Duration100Comp}
	\end{figure*}

	\begin{figure*}[!htb]
		\centering
		\hspace*{\fill}%
		\begin{subfigure}[t]{.45\linewidth}
			\includegraphics[width=\linewidth, height=.45\linewidth]{graphs/tikz-slices/NA_025_Density_slice_0_um_100_nJ_100_fs_100_um.tikz}
			\caption{$ z=\SI{100}{\micro\meter} $, $ \text{NA} = \num{0.25} $}
			\label{subfig:SliceFCDensity025100100Comp}
		\end{subfigure}\hfill%
		\begin{subfigure}[t]{.45\linewidth}
			\includegraphics[width=\linewidth, height=.45\linewidth]{graphs/tikz-slices/NA_085_Density_slice_0_um_100_nJ_100_fs_100_um.tikz}
			\caption{$ z=\SI{100}{\micro\meter} $, $ \text{NA} = \num{0.85} $}
			\label{subfig:SliceFCDensity085100100Comp}
		\end{subfigure}\hspace*{\fill}%
		
		\hspace*{\fill}%
		\begin{subfigure}[t]{.45\linewidth}
			\includegraphics[width=\linewidth, height=.45\linewidth]{graphs/tikz-slices/NA_025_Density_slice_0_um_100_nJ_100_fs_250_um.tikz}
			\caption{$ z=\SI{250}{\micro\meter} $, $ \text{NA} = \num{0.25} $}
			\label{subfig:SliceFCDensity025100250Comp}
		\end{subfigure}\hfill%
		\begin{subfigure}[t]{.45\linewidth}
			\includegraphics[width=\linewidth, height=.45\linewidth]{graphs/tikz-slices/NA_085_Density_slice_0_um_100_nJ_100_fs_250_um.tikz}
			\caption{$ z=\SI{100}{\micro\meter} $, $ \text{NA} = \num{0.85} $}
			\label{subfig:SliceFCDensity085100250Comp}
		\end{subfigure}\hspace*{\fill}%
		\caption{Slice of the generated free carrier density for $ t_0=\SI{100}{\femto\second} $, $ f_\text{inside}=\SI{100}{\micro\meter} $/$f_\text{inside} =\SI{250}{\micro\meter} $, $ E_0=\SI{100}{\nano\joule} $ focused with a lens with an NA = \num{0.25}/\num{0.31}/\num{0.35}/\num{0.41} in \Cref{subfig:SliceIntensity025100100Comp,subfig:SliceIntensity025100250Comp} and an NA = \num{0.85}/\num{1.07}/\num{1.18}/\num{1.40} in \Cref{subfig:SliceIntensity085100100Comp,subfig:SliceIntensity085100250Comp} directly at the focal spot ($ z=\SI{100}{\micro\meter} $/$ z=\SI{250}{\micro\meter} $). }
		\label{fig:SlicesNA085Depth100Energy100Duration100CompDensity}
	\end{figure*}

	\bibliography{bibliography_clean.bib}
\end{document}